\documentclass{aa}
\usepackage{graphicx}
\usepackage{txfonts}

\usepackage{diagbox}
\usepackage{rotating}
\usepackage{tikz}
\usepackage{multirow}
\usepackage{wasysym}
\usepackage{adjustbox}
\usepackage{afterpage}
\usepackage{placeins}
\usepackage{xcolor}
\usepackage{caption}
\usepackage[strict]{changepage}
\usepackage{lscape}
\usepackage{pdflscape}
\usepackage{soul}

\usepackage{hyperref}
\hypersetup{colorlinks=True,citecolor=blue}

\def\ms{\hbox{\,m\,s$^{-1}$} }         
\def\cms{\hbox{\,cm\,s$^{-1}$} }       
\def\m2s2{\hbox{\,m$^{2}$\,s$^{-2}$} } 
\def\vsini{\hbox{$v$\,sin\,$i$}}      
\def\sini{\hbox{sin\,$i$}}      
\def\Mearth{\hbox{$\mathrm{M}_{\oplus}$}}             
\def\Msun{\hbox{$\mathrm{M}_{\odot}$}}             

\def\ang{\text{\AA}}

\usepackage{amstext}

\begin{document}

    \title{YARARA V2: Reaching sub\,\ms{} precision over a decade using PCA on line-by-line RVs}


   \author{M. Cretignier\inst{1,2}
          \and X. Dumusque \inst{1}
          \and S. Aigrain \inst{2}
          \and F. Pepe \inst{1}
          }

   \institute{
   $^1$ Astronomy Department of the University of Geneva, 51 ch. de Pegasi, 1290 Versoix, Switzerland\\
   $^2$ Department of Physics, University of Oxford, OX13RH Oxford, UK \\ 
            }

   \date{Received XXX ; accepted XXX}

 
  \abstract
   {The detection of Earth-like planets with the radial-velocity (RV) method is extremely challenging today due to the presence of non-Doppler signatures such as stellar activity and instrumental signals that mimic and hide the signals of exoplanets. In a previous paper, we presented the YARARA pipeline, which implements corrections for telluric absorption, stellar activity and instrumental systematics at the spectral level, then extracts line-by-line (LBL) RVs with significantly better precision than standard pipelines.}
   {In this paper, we demonstrate that further gains in RVs precision can be achieved by performing Principal Component Analysis (PCA) decomposition on the LBL RVs.}
   {The mean-insensitive nature of PCA means that it is unaffected by true Doppler shifts, and thus can be used to isolate and correct nuisance signals other than planets.}
   {We analysed the data of 20 intensively observed HARPS targets by applying our PCA approach on the LBL RVs obtained by YARARA. The first principal components show similarities across most of the stars and correspond to newly identified instrumental systematics, which we can now correct for. For several targets, this results in an unprecedented RV root-mean-square of around 90\,\cms{} over the full lifetime of HARPS. We use the corrected RVs to confirm a previously published 120-day signal around 61\,Vir, and to detect a Super-Earth candidate ($K\sim60\pm6$\,\cms{}, $m$ \sini{} = $6.6\pm0.7$\,\Mearth{}) around the G6V star HD20794, which spends part of its 600-day orbit within the habitable zone of the host star.}
   {This study highlights the potential of LBL PCA to identify and correct hitherto unknown, long-term instrumental effects and thereby extend the sensitivity of existing and future instruments towards the Earth analogue regime.}


   \maketitle

\section{Introduction}

The detection of Earth-like exoplanets orbiting Sun-like stars remains one of the most exciting perspectives for the future of astrophysics, but also one of the most tremendous challenges for the next few years. Until now, such detections have been out of the reach of the radial velocity (RV) technique, as the most precise spectrographs: HIRES \citep{Vogt(1994)}, HARPS \citep{Mayor(2003)} and HARPS-N \citep{Cosentino(2012)} typically reached a precision of $\sim$1\,\ms{}. This is an order of magnitude larger than the 0.1\,\ms{} RV semi-amplitude that the Earth induces on the Sun.

Despite the technical challenges involved in reach the extreme precision required, the RV method remains the most promising technique for the detection of other Earths around stars closer than 15 pc, at least in the next decade, due to the low transit probability of these objects and the extremely dim light emitted or reflected by their surface \citep{Zhu(2021)}. This has motivated the design of a new generation of ultra-stable spectrographs such as ESPRESSO \citep[][]{Pepe:2020aa}, EXPRES \citep[][]{Jurgenson:2016aa} and NEID \citep{Schwab(2016)} that have already demonstrated an RV precision of $\sim$50\,\cms on a timescale of a few months \citep[][]{Suarez-Mascareno:2020aa,Brewer:2020aa,Lin(2022)}. While this represents a major step toward achieving the necessary sensitivity, it remains to be seen whether this level of precision is sustained over multi-year timescales.

In recent years, significant progress has been made in improving the correction of telluric, instrumental and stellar effects at the level of spectra, line profiles and/or RV time-series \citep[see][and references therein]{Zhao(2022)}. In particular, the YARARA pipeline \citep{Cretignier(2021)} corrects for these effects at the spectral level, then extracts line-by-line (LBL) RVs that can be combined into a global RV with significantly improved precision compared to the HARPS Data Reduction Software (DRS v3.5). 
The present paper seeks to improve on this further by applying Principal Component Analysis (PCA) on the LBL RVs to identify, isolate and correct non-Doppler signals that have persisted through the spectrum-level post-processing.

The remainder of this paper is structured as follows. Our methodology is described in Sect.~\ref{sec:method}. Specifically, Sect.~\ref{sec:processing} describes the pre-processing of the data, Sect.~\ref{sec:PCA} presents the mathematical framework for PCA correction of LBL RVs, Sects.~\ref{sec:snr} and \ref{sec:folding}, introduce strategies for boosting the signal-to-noise of individual components. Then, in Sect.~\ref{sec:systematics} we use the most significant components to calibrate and correct for newly identified systematic effects, while Sect.~\ref{sec:iterations} summarize the overall cascade reduction. Finally, Sect.~\ref{sec:keplerians} describes how we explicitly include Keplerian signals into the model. 
In Sect.~\ref{sec:results}, we then apply this methodology to five targets intensively observed by HARPS. In the case of HD10700 (Sect.~\ref{sec:hd10700}), we perform an injection-recovery test to show that the method preserves planetary signals. We then turn to HD192310 (Sect.~\ref{sec:hd192310}), which presents a clear stellar rotation signal, and to HD115617 (Sect.~\ref{sec:hd115617}), where we confirm a previously detected planet candidate. HD109200 (Sect.~\ref{sec:hd109200}) is a complex case, which remains unresolved, and illustrates the limitations of our method when the signal-to-noise ratio (S/N) is lower than 200. Finally, we analyse HD20794 (Sect.~\ref{sec:hd20794}), where we detect a new candidate exoplanet. We then conclude in Sect.~\ref{sec:conclusion}.

\section{Method}
\label{sec:method}

The present section describes the global LBL PCA framework used to improve RV precision. In particular, the mathematical framework illustrating how PCA can disentangle planetary signals from systematics is presented. The limitations of the method itself are addressed and we proposed some solutions to counteracts them.

\subsection{Data pre-processing
\label{sec:processing}}

 We worked with HARPS 1D-merged spectra produced by the official data reduction software (DRS). Only spectra after BJD = 2453500 were processed since RV time-series before that date obtained on standard quiet stars show unusual RV excursion values due to commissioning (priv. comm.). Also, only spectra before the fiber upgrade of the instrument in 2015 \citep{LoCurto(2015)} were considered. Despite an expected improvement in instrumental stability with the new fibers, the RV precision is worse after the fiber upgrade on several standard stars (see for instance \citealt{Cretignier(2021)}), because the data reduction was not optimised for this new version of the instrument\footnote{We note that this issue has been solved by version 3.0.0 of the ESPRESSO pipeline that has recently been optimized for HARPS in the same manner as was done for HARPS-N by \protect\citet{Dumusque(2021)}. While this new version of the pipeline is publicly available at \protect\url{https://www.eso.org/sci/software/pipelines/espresso/espresso-pipe-recipes.html}, it has now yet been fully validated, and was therefore not used in the present analysis.}. Moreover, after the fiber upgrade, the spectrograph should be considered as a new instrument and not enough public observations are available to properly apply our post-processing method with this limited dataset at the moment. 
 
 Where multiple observations of a given star were taken within a given night, the corresponding 1D-merged spectra were stacked. All nightly-stacked, 1-D spectra were then continuum-normalised using RASSINE \citep{Cretignier(2020b)}, then post-processed using the YARARA pipeline \citep{Cretignier(2021)} to remove known systematics present in HARPS spectra (instrumental and telluric contamination) at the spectrum level. In order of processing, YARARA corrects for i) cosmics, ii) tellurics, iii) an interference pattern that is present on the detector, iv) stellar activity, v) point spread function variation, vi) ghosts, vii) stitching between different sub-arrays of the detector and viii) contamination from the simultaneous calibration fibre (fibre B). To derive more accurate LBL RVs \citep{Dumusque(2018)}, a data-driven line selection was performed for each star, following \citet{Cretignier(2020a)}. Since some residual stellar activity signals were shown to survive, we also correct LBL RVs using the "shell" methodology developed by \citet{Cretignier(2022)}. 

 As a reminder, the shell method is a variant of template matching methods (see e.g \citet{Zechmeister(2018),Silva(2022)}), that aim to measure the RV shifts compared to a reference spectrum $S_{ref}(\lambda)$. Rather than measuring the shift, the shell method extracts the lines profile distortions that are orthogonal to a pure Doppler shift, and their associated time-domain coefficients. In practice, the reference spectrum $S_{ref}(\lambda)$ is taken as the median of the YARARA-corrected spectra, shifted according to the current $\text{RV}(t)$ measurements. The difference  $\delta(\lambda)=S_{obs}(\lambda) - S_{ref}(\lambda)$ between any observed spectra $S_{obs}(\lambda)$ and the master spectrum can be expressed as a function of the master spectrum itself and its wavelength derivative $\delta(S_{ref},\partial S_{ref} / \partial \lambda)$. Projecting the observed spectrum within that bounded space, then allows us to fit simultaneously for the Doppler shift and some line profile distortions that affect lines of different depths in different ways. 
 
 The next section explains how LBL RVs can be further corrected for systematics thanks to PCA. Note that at any time, a given line selection containing $l$ stellar lines can be combined by weighted averaging, to produce a single RV time-series (see \citealt{Dumusque(2018)}), where the weights $\omega_i$ are defined as the inverse squared RV uncertainties: 
\begin{equation}
\label{eq:1}
\text{RV}(t) = \sum^l_{i=1}\omega_i \cdot \text{RV}_i(t) = \sum^l_{i=1}  \left( \frac{1}{\sigma_{\text{RV}_i}(t)^2} \right) \cdot \text{RV}_i(t).
\end{equation}

Since outliers could bias the PCA afterwards, we rejected the 5\% of the stellar lines with the largest relative dispersion, defined as the ratio of the standard deviation of the RV time-series with the median of the RV uncertainties.
 
\subsection{Application of PCA to LBL RVs to detect non-Doppler signatures
\label{sec:PCA}}

The LBL RV signal of an individual stellar line $\text{RV}_i(t)$ can always be described as the superposition of several components. A component that affects all the stellar lines in a similar way will be called Doppler and those affecting each stellar line differently will be called non-Doppler. The RV signals of exoplanets $\text{RV}_p(t)$ belong to the first category, whereas stellar activity and instrumental signals belong to the second. However, even a non-Doppler effect can mimic a true Doppler shift in the sense that it can shift the spectrum as a whole. As already discussed in \citet{Cretignier(2022)}, this explains why methods which separate pure Doppler shifts from distortions of the mean line profile do not result in a perfect correction of these effects.

For a system containing $n$ planets, each with Keplerian signal $\text{K}_p(t)$ ($p=1,\ldots,n$), and $N$ non-Doppler effects, each with time-dependence $\text{V}_{j}(t)$ ($j=1,\ldots,N$), the RV of $i^{\rm th}$ stellar line at time $t$ is given by:  
\begin{equation}
\label{eq:3}
\begin{array}{rcccl}
\text{RV}_i(t) & = & \displaystyle\sum^n_{p=1}\text{K}_p(t) & + & \displaystyle\sum^N_{j=1} a_{i,j}\cdot \text{V}_{j} (t) \\[2pt]
& \equiv & \text{RV}_P(t) & + & \displaystyle\sum^N_{j=1} a_{i,j}\cdot \mathrm{V}_{j} (t),
\end{array}
\end{equation}
where $a_{i,j}$ determines how strongly the $i^{\rm th}$ line is affected by the $j^{\rm th}$ non-Doppler signal.

\begin{figure}[t]
	
	\centering
	\includegraphics[width=9cm]{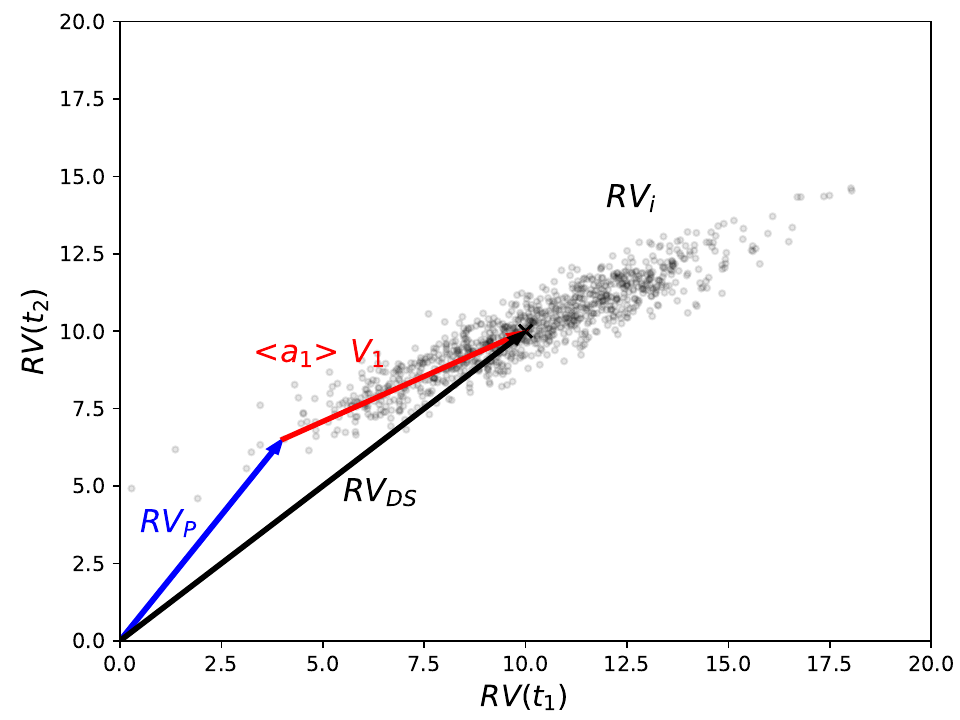}
	\caption{
	Schematic representation of PCA applied on LBL RVs. The number of dimensions of the space is equal to the number of observations (here two for ease of visualisation), while each element of the data cloud is made of one LBL RV time-series $\text{RV}_i(t)$. The Doppler shift $\text{RV}_{DS}$ (black arrow), which defines the barycenter of the cloud, is equal to a planetary signal $\text{RV}_P$ (blue arrow) plus a mean systematic effect $\text{V}_1$ (red arrow). Because PCA is mean-invariant, the first PC will be along the $\text{V}_1(t)$ direction and can be used to correct the $\text{RV}_i(t)$. }
	\label{FigPCA}
	
\end{figure} 

If the weighted average  $<a_{j}> = \sum^l_{i=1}\omega_i \cdot a_{i,j} \neq 0$, a non-Doppler effect can induce a net "true Doppler shift". To make this explicit, we decompose the coefficients into a "mean" and a "variance" term: $a_{i,j} = <a_{j}> + b_{i,j}$, where $<b_{j}> = \sum^l_{i=1}\omega_i \cdot b_{i,j} = 0 $ by construction. Substituting this into Eq.\ref{eq:3}: 
\begin{equation}
\label{eq:5}
\begin{array}{rcccl}
\text{RV}_i(t) & = & \text{RV}_P(t) + \displaystyle\sum^N_{j=1} <a_{j}> \cdot  \text{V}_{j}(t) & + &  \displaystyle\sum^N_{j=1} b_{i,j}\cdot \text{V}_{j} (t) \\[2pt]
& \equiv & \text{RV}_{DS}(t) & + &  \displaystyle\sum^N_{j=1} b_{i,j}\cdot \text{V}_{j} (t),
\end{array}
\end{equation}
we can see that the total Doppler-like signal $ \text{RV}_{DS}(t)$ is the sum of the combined planetary signal and the mean component of the non-Doppler signals. 

In this work, we use PCA using the Python package \textit{Scikit-learn} \citep{Pedregosa(2011),Grisel(2021)} to decompose a time-series of LBL RVs into orthogonal components. The first step in any PCA procedure is to "center" the rows of the input matrix such that they have zero mean: $\widehat{\text{RV}}_i(t) = \text{RV}_i(t)\, - <\text{RV}>(t)$. The extracted Principal Components (PCs) can thus be expected to trace only non-Doppler signals and should be insensitive to planetary signals\footnote{This is true if the mean RV value is unbiased which could require RV uncertainties (for the weighted average) that accurately measure the intrinsic RV accuracy. In our case, we did not found any significant difference or advantage to include the LBL RV uncertainties as weights for the PCA. This could be explained, because low SNR observations or anomalous spectra have been rejected by YARARA, YARARA has already cleaned most of the local systematics and a sigma-clipping has been performed on the lines. Also, PCA will mostly be applied on averaged LBL RVs (see next section).} (as well as to the Doppler-like component of non-Doppler signals):
\begin{equation}
\label{eq:4}
\widehat{\text{RV}}_i(t) = \sum^N_{j=1} \beta_{i,j}\cdot \text{PC}_{j} (t) = \sum^N_{j=1} b_{i,j}\cdot \text{V}_{j} (t) 
\end{equation}
where $\beta_{i,j}$ is the coefficient, or score, linking the $j^{\rm th}$ principal component $\text{PC}_{j} (t)$ to the $i^{\rm th}$ line, and we have used the fact that $<\text{RV}>(t) \equiv \text{RV}_{DS} (t)$. 

This decomposition is illustrated in Fig.~\ref{FigPCA}, which schematically shows LBL RVs for 1000 lines at two times $t_1$ and $t_2$. The first PC defines the direction $\text{V}_1(t)$, which represents the main axis of variance of the 2-D cloud. 
Directly subtracting the component of the individual LBL RVs that can be explained by the first PC would not enable us to correct for the "mean" systematic effect, represented on Fig.~\ref{FigPCA} by the red arrow. 

To correct for the systematic effects in the original LBL RVs (not mean-subtracted), we "de-project" them onto the PCs. In other words, we use the $\text{PC}_j(t)$ as a substitute for the $\text{V}_j(t)$ in Eq.~\ref{eq:3}: 

\begin{equation}
\label{eq:11}
\text{RV}_i(t) = \text{RV}_P(t) + \sum^N_{j=1} \alpha_{i,j}\cdot \text{PC}_{j} (t)
\end{equation}

and then fit for the coefficients $\alpha_{i,j}$.This allows us to correct not only for the variance, but also (at least in part) for the mean component of the systematic effects. In the example shown in Fig.~\ref{FigPCA}, doing so would also absorb much of the planetary signal, but this is because the space is of low dimensionality (only 2 observations) and the first PC is almost collinear with the planetary Doppler shift between the two epochs. In a realistic case, with many more observations, and hence more dimensions, such effects are much reduced and can be assessed \citep{Cretignier(2022)}. 

The present method is somewhat similar to the one presented in \citet{Cretignier(2022)} except that the PCA is now performed on the LBL RVs (rather than on the spectra). For that reason, the same caveats apply: 
\begin{enumerate}
    \item Large cross-term correlation may exist between PCs and the planetary signal $\text{RV}_P$;
    \item PCA is affected by outliers and low S/N of the data;
    \item as real-world nuisance signals are not necessarily mutually orthogonal, each PC often contains a mixture of several physical effects, which can be difficult to disentangle or interpret ($\text{PC}_j$ $\neq$ $\text{V}_j$ in Eq.\ref{eq:4});
    \item the ordering of the PCs and the mixture of physical effects within each of them changes from star to star;
    \item the largest directions of variance are not necessarily the axes with the largest mean (so with the largest RVs effect);
\end{enumerate}

Item 1 mostly affects the detectability of planetary signals with periods that are long compared to the observational baseline, where chance of cross-term with a nuisance signal is most likely to arise, or specific periods such as 1-year and its harmonics. However, this problem can be mitigated by explicitly including the Keplerian signals alongside the PCs in the final fit (see Sect.~\ref{sec:iterations}). The remainder of this section describes strategies we developed to address the other items.


\subsection{Improving the PCA decomposition
\label{sec:snr}}

Item 2 is related to noisy datasets. The precision of the LBL RVs directly affects the precision of the PCs we extract from them. For stars with moderate S/N ($S/N_{cont} \sim 200$), the uncertainty on individual LBL RVs is usually around 15\,\ms{}. Even for the brightest stars ($S/N_{cont} \sim 600$), in never drops below 5\,\ms{}. This is one order of magnitude larger than the mean amplitudes of the systematics that we are trying to correct, which are below the\,\ms{} level \citep{Cretignier(2021)}. Moreover, the presence of outliers is concerning since the PCA will try to capture their behaviour (due to their large variance) whereas their final mean effect is negligible. For example, in HARPS data, LBL PCA revealed a new interference pattern (see Appendix \ref{appendix:b}), with an amplitude smaller than 0.1\% in flux, which strongly affects LBL RVs of stellar lines in the blue, but its final effect on the global RVs is smaller than 10\,\cms{}. This simple example case led us to conclude that we should not apply PCA \textit{solely} on individual LBL $\text{RV}_i$. 

\begin{figure}[t]
	\centering
	\includegraphics[width=9cm]{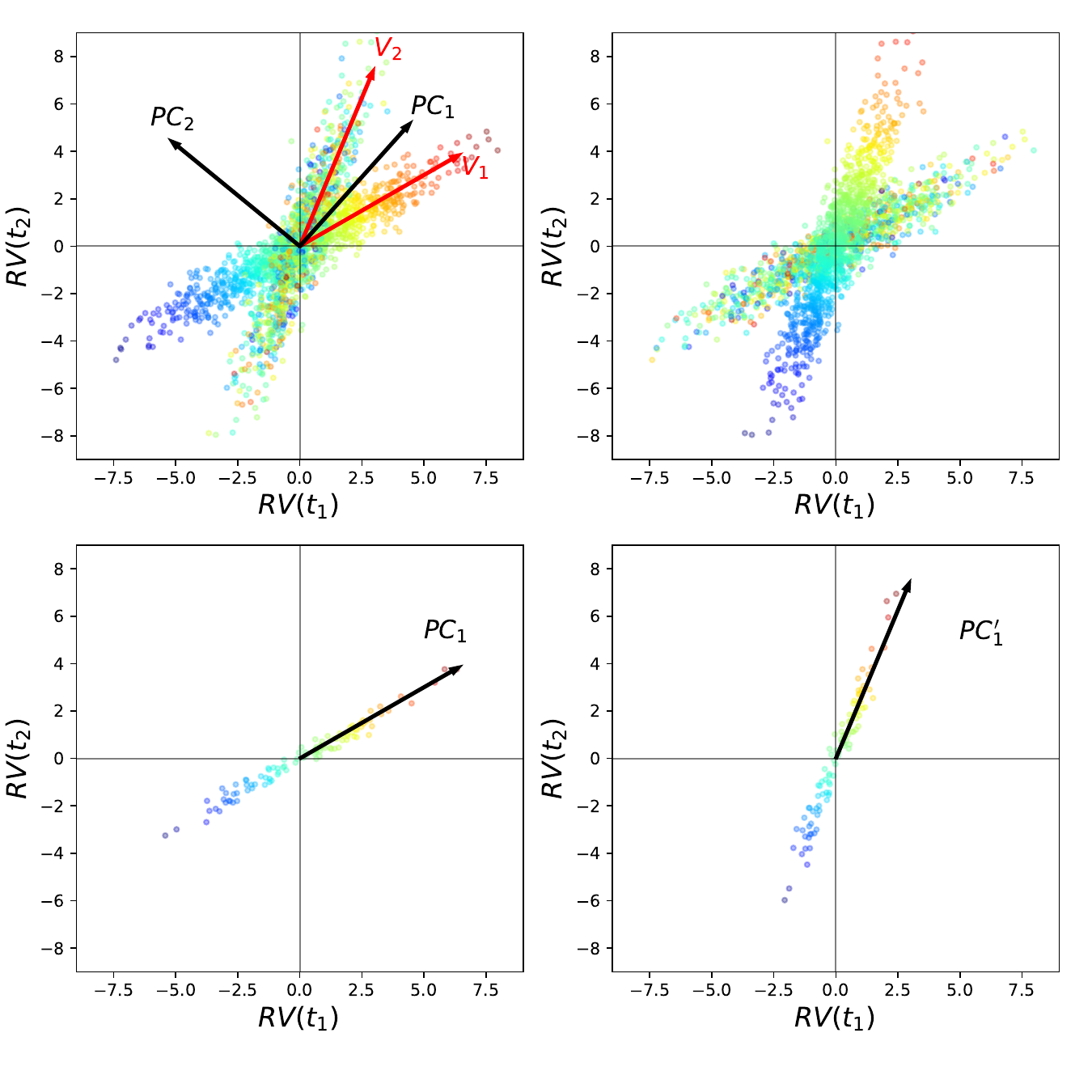}
	\caption{Same as Fig.~\ref{FigPCA} for a mixture of two systematic effects sharing some partial correlation. 
 The individual LBL RVs are shown in the top row, coloured according to wavelength (left) or y-pixel position (right). The two systematic effects $V_{1}$ and $V_{2}$ (shown by the red arrows in the top left panel) depend, respectively on these two variables. A classical PCA decomposition results in orthogonal components $\text{PC}_1$ and $\text{PC}_2$ (as shown by the arrows in the top left panel), that are unrelated to the main systematics effects. In the bottom row, we first group-average the LBL RVs according to wavelength (left) and y-position (right) before performing PCA on the result to identify principal components $\text{PC}_1$ and $\text{PC}^\prime_1$ that are proportional to $\text{V}_{1}$ and $\text{V}_{2}$, respectively. The same would occur for a cloud of data with outliers.}
	\label{FigPCA2}
	
\end{figure} 

The most trivial solution therefore consists in stacking or averaging LBL RVs in order to boost the S/N, where the weighted average of a selection of lines is computed with the inverse squared of the LBL $\text{RV}_i$ uncertainties as in Eq.\ref{eq:1}. In fact, most of the issues listed in items 2--5 (varying mixtures of contamination, orthogonality, noise and outliers) can be solved by averaging over appropriately selected groups of lines. The only question is then how to define these groups of lines. The answer to that question depends on the effects that we want to correct for. 

Let us take an example to illustrate the idea, which we illustrate schematically in Fig.~\ref{FigPCA2}. Let us assume that a star is affected simultaneously by two different types instrumental systematics, with mean-effects $\text{V}_{1}$ and $\text{V}_{2}$. In this example, the first effect is related to a flux anomaly that affects blue spectral lines more strongly than red ones, whereas the second affects some specific pixel columns in the original 2-D images. In this simplistic example, each of the simulated stellar lines is affected by one or the other, but not both. In reality, each line would be affected by a mixture of effects, but the point we are trying to make here is that it is possible to identify groups of lines whose mean RVs are predominantly affected by a given type of systematics. As shown in Fig.~\ref{FigPCA2}, a direct PCA decomposition of the LBL RVs will result in PCs that are a mixture of the two systematic effects ($\text{PC}_j$ $\neq$ $\text{V}_j$ in Eq.\ref{eq:4}). However, grouping lines in large wavelength chunks will strongly mitigate the signal of the column signatures, while allowing the PCA to isolate the wavelength-dependent effect. On the other hand, group lines by y-pixel position on the detector will strengthen the effect of the bad pixel columns. This can be used to identify PCs that are more directly linked to the individual systematic effects, as shown in the bottom row of Fig.~\ref{FigPCA2}. This example demonstrates that line averaging can be particularly powerful, provided that one can identify a suitable way to create the groups.     

In the absence of a detailed \emph{a priori} understanding of the origin of the systematics, we can identify some parameters that are well-defined and most likely to control the extent to which a given systematic effect affects a given line. The most obvious such parameter is, naturally, the wavelength $\lambda$, since both instrumental and stellar activity effects are expected to produce chromatic variations \citep{Coffinet(2019), Cersullo(2019), Zechmeister(2020)}. We therefore selected this parameter to drive our line selections.

\subsection{Detecting chromatic effects with chunk-by-chunk (CBC) RVs
\label{sec:folding}}

\begin{figure*}
	
	\centering
	\includegraphics[width=18cm]{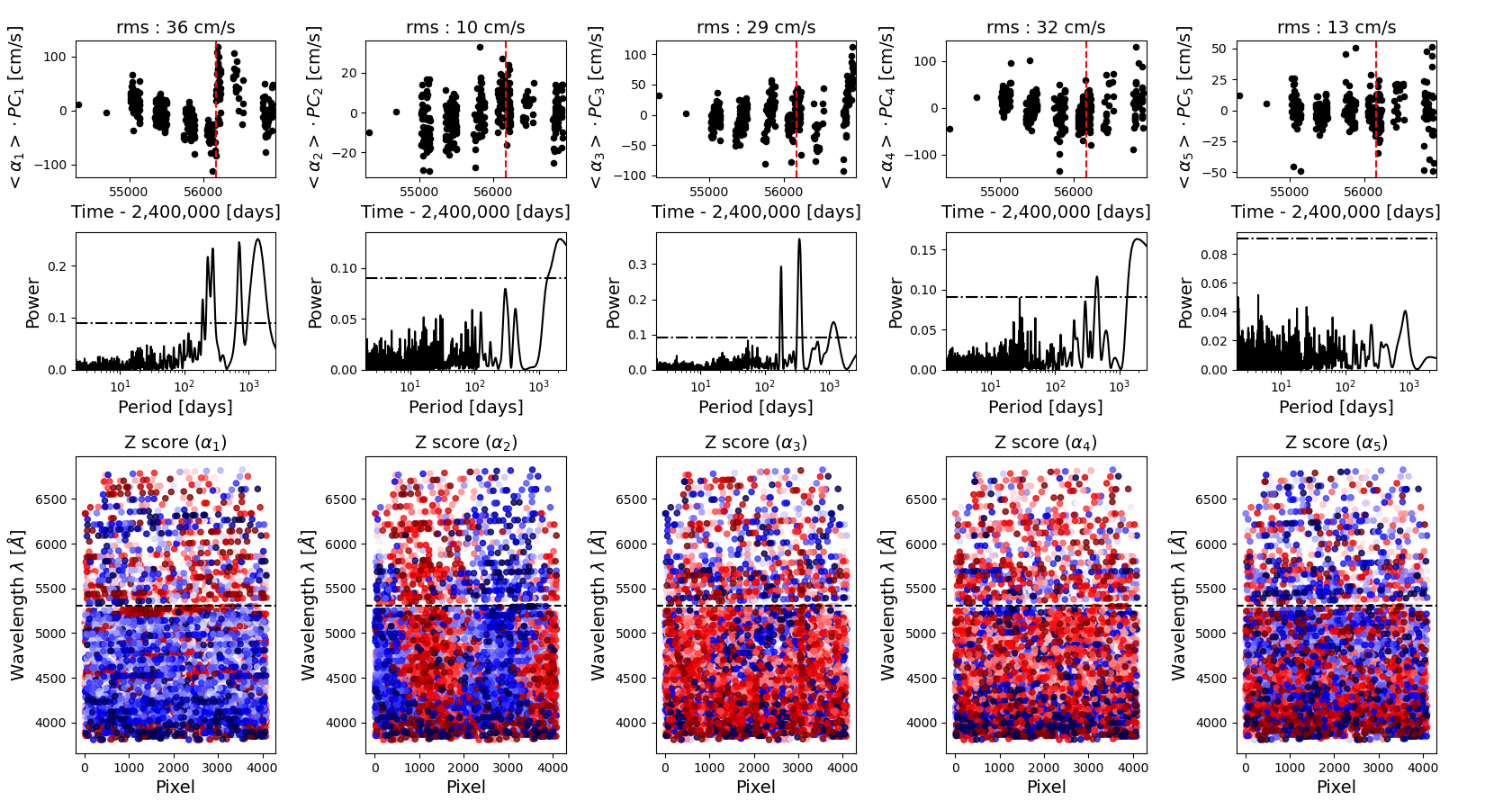}
	\caption{
	Representation of the five first PCs obtained on the chunk-by-chunk RVs of HD192310. \textbf{Top:} $\text{PC}_j(t)$ length-projected onto the RV(t) time-series. The date(s) of ThAr lamp replacement(s) are displayed as vertical red dashed line(s). \textbf{Middle:} Corresponding Generalised Least Squares (GLS) periodogram of the PCs. \textbf{Bottom:} Projection of the $\alpha_{i,j}$ coefficients (converted to $Z$-scores, i.e.\ normalised to zero mean and unit variance) into the physical detector space (pixel, $\lambda$). The separation between the two detectors of HARPS around $\lambda=5250 \AA$ is indicated by the horizontal dashed line. The color scale was set between $Z=-2$ and $Z=2$, the direction being irrelevant since signs of the PCs are free to change. $\text{PC}_1(t)$ is clearly related to ThAr lamp ageing as highlighted by the discontinuity of 100\,\cms{} visible in the time-domain that matches the date of the lamp's replacement. $\text{PC}_2(t)$ shows some smooth modulation in the physical detector space, whereas $\text{PC}_3(t)$ and $\text{PC}_4(t)$ exhibit power at a period of 1-year in the GLS.}
	\label{FigColor1}

	\centering
	\includegraphics[width=18cm]{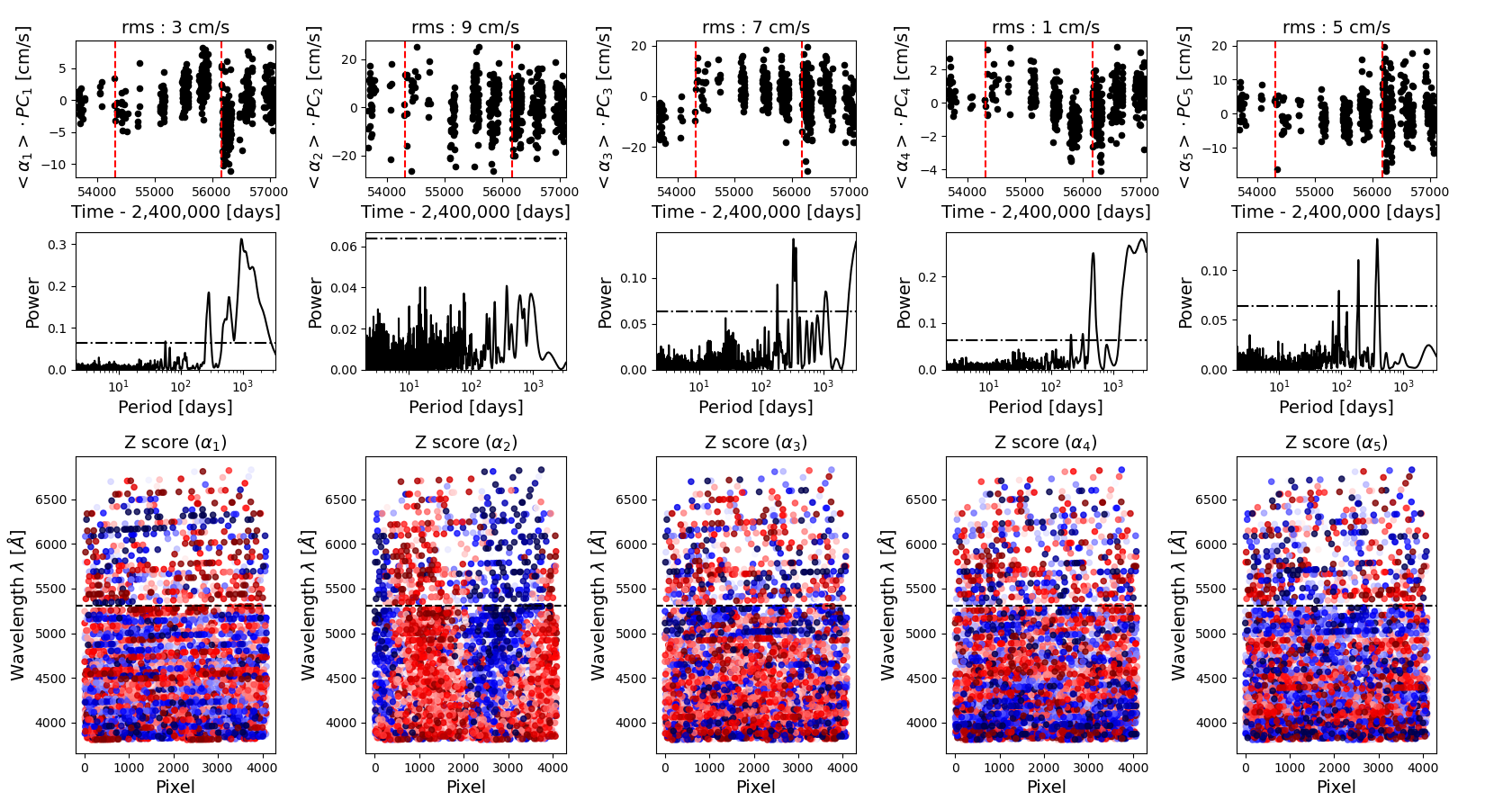}
	\caption{
	Same as Fig.~\ref{FigColor1}, but for HD20794. The first two components, $\text{PC}_1(t)$ and $\text{PC}_2(t)$, display a qualitatively similar behaviour (and hence origin) as for HD192310. However, we note that the sign of the offsets due to ThAr ageing on $\text{PC}_1(t)$ is now reversed and smaller with a jump of 10\,\cms{}. The three components $\text{PC}_3(t)$, $\text{PC}_4(t)$ and $\text{PC}_5(t)$ all present 1-year power. For this star, the rms of each of the components fit to the RVs according to Eq.\ref{eq:11} ($<\alpha_j>\cdot \text{PC}_j(t)$) is smaller than 10\,\cms{}.}
	\label{FigColor2}
	
\end{figure*}

As shown in the previous Sect.~\ref{sec:snr}, the correction of LBL RVs by PCA can be improved if the latter are averaged using some parameters that correlate with the direction of variance of the systematics (as in Fig.~\ref{FigPCA2}), and wavelength is a natural parameter to use for this.
We thus averaged the LBL RVs over 4 \ang{} "chunks", resulting in "chunk-by-chunk" (CBC) RVs. The first chunk was defined to start at the wavelength of the bluest first stellar line. We tested a range of chunk width, ultimately settling on 4 \ang{} as the best trade-off between S/N improvement, mitigation of the interference pattern signature (which has a periodicity of 0.1 \ang{}, see Appendix \ref{appendix:b}) and sensitivity to smooth chromatic trends in the LBL RVs. 
We performed PCA on the CBC RVs and examined the behaviour of the first five components for our targets. An example for the star HD192310 is displayed in Fig.~\ref{FigColor1}. Note that HD192310 contains a clear planetary signal of semi-amplitude 2.5\,\ms{} at 75 days (see Sect.~\ref{sec:hd192310}), whereas such signal is absent from our PCs, clearly demonstrating that PCA is insensitive to planetary signals, which are mean-effects. 

We identified two components that are common to most of the HARPS stars we reduced, indicating a clear instrumental origin (see Fig.~\ref{FigColor1} and Fig.~\ref{FigColor2}). Instrumental systematics are expected to dominate on HARPS, since the observed stars were rather quiet, and we had already applied YARARA and the shell framework, which should correct part of the stellar activity. 

We confirmed the instrumental origin of the trends identified in the CBC RVs by examining the $\text{PC}_j(t)$ time-series. The first, $\text{PC}_1(t)$, displays a discontinuous behaviour, with jumps corresponding to the dates at which the ThAr lamp was replaced. This indicates that the change in the
RV zero-point of the instrument caused by the lamp replacement is not fully corrected by version 3.5 of the DRS. This phenomenon was also detected for HARPS-N, and corrected in the newer DRS version (version 2.2.3, \citealt{Dumusque(2021)}. The slow drift between the offsets is explained by the slow ageing of the lamp that modifies the internal pressure, changing the positions of the reference ThAr lines on the detector. 

The second component, $\text{PC}_2(t)$, does not present such a clear time-domain signature, nor specific periodicity, but is rather noisy. In order to confirm that this effect is clearly instrumental, we projected back the $\alpha_{i,j}$ coefficients onto the detector space ($x, \lambda$), where we use the stellar line wavelength $\lambda$ to trace the cross-dispersion direction, and the $x$-pixel coordinate varies along the dispersion direction within each order. In that space, the second PC score displays a smooth structure with left-to-right variation, which is clearly due to an instrumental systematic effect, although its origin is not well understood at the moment. One possible explanation may be an imperfect fiber injection due to guiding errors and seeing variation which would explain why no clear periodicity is found since both guiding and seeing are stochastic processes. If so that component should be missing when analysing HARPS data post 2015, since new octagonal fibres were installed with a better scrambling of the light input injection. 

\begin{figure*}
	
	\centering
	\includegraphics[width=18cm]{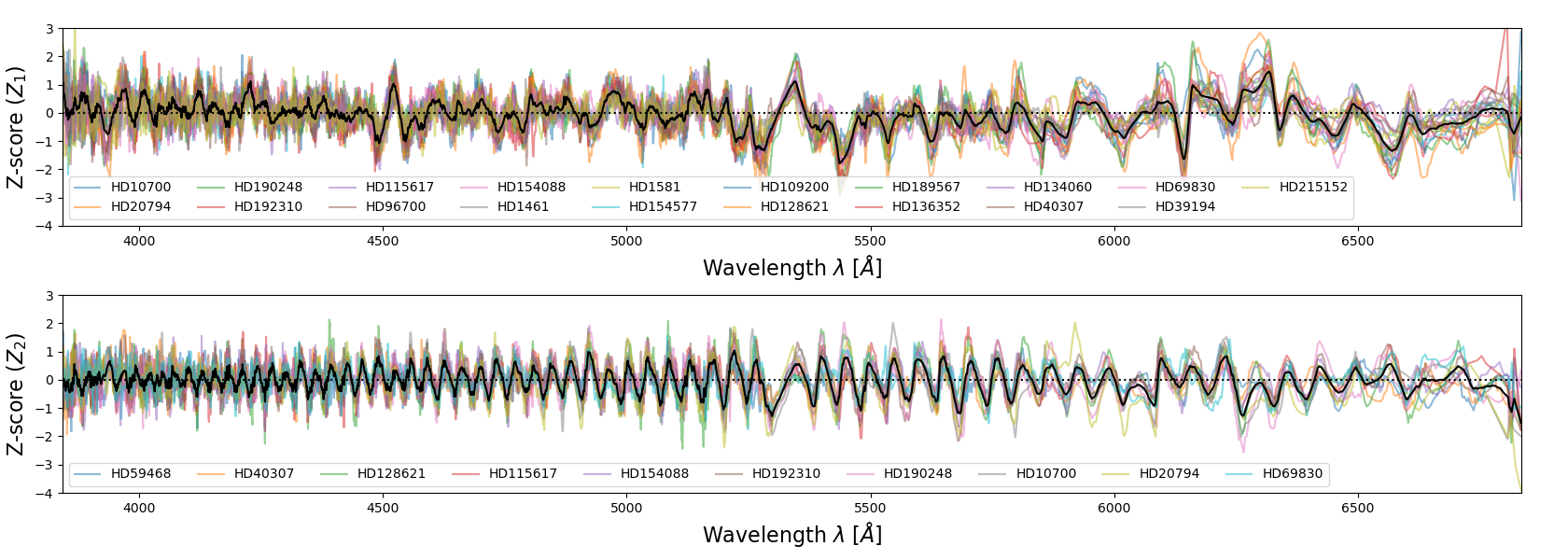}
	\caption{
	$Z_j(\lambda)$ calibrations curves (black curves) obtained by stacking the Z-score converted $\beta_{i,j}$ coefficients of several HARPS targets (name in the labels). \textbf{Top:} $Z_1(\lambda)$ curve related to the ThAr ageing and lamp offsets correction. \textbf{Bottom:} $Z_2(\lambda)$ curve related to an unknown instrumental effect with a smoothed structure across the physical detector.}
	\label{FigSlice1}

	
\end{figure*} 

An important observation to raise is that, despite a clearly similar origin for the components, the time-series of both stars are very different. This could be surprising or "contre-intuitive" when thinking about the first PC related to ThAr ageing and wavelength solution. Indeed, whereas a positive offset of 100\,\cms{} is observed on HD192310 (Fig.~\ref{FigColor1}) for the first component $\text{PC}_1(t)$, a negative offset of 10\,\cms{} is measured for HD20794 (Fig.~\ref{FigColor2}). This example demonstrates that instrumental systematics cannot be simply averaged in the time-domain to form some sort of "master time-vector" with which to perform corrections \citep{Trifonov(2020)}. This is not altogether surprising, given that each star probes the detector space ($x, \lambda$) differently, depending on its peculiar systemic radial velocity, spectral type and/or metallicity, resulting in different mean-effects in the final LBL RVs. A similar observation was already made by \citet{Cretignier(2021)} in the spectral domain to explain why some stars where more sensitive to the "detector stitching" effect than others. In the next section, we describe how we can exploit the common nature of the systematic trends we have identified in the CBC RVs while adapting the correction to the individual behaviour of each star.

\subsection{Correction of the instrumental systematics 
\label{sec:systematics}}

By looking at the five first PCs of dozens of HARPS targets, we found that several types of trends recur across many stars. 
This observation can be exploited to better constrain the correction of the systematics by reducing the freedom given to the PCA. 

If we were able to average the LBL RVs according to the strength coefficient $b_{i,j}$ of a specific $\text{V}_j(t)$ contamination (see Eq.\ref{eq:4}), our problem would be solved. However, we never know the $b_{i,j}$ coefficients \emph{a priori}. We only have access to the $\beta_{i,j}$ coefficients and the $\text{PC}_j(t)$. We noted that, once converted to $Z$-scores\footnote{The $Z-$score of a variable $X$ drawn from a distribution is defined as the number of standard deviations away from the mean. In other words, converting a variable to a $Z$-score consists in subtracting the sample mean $\mu$ and dividing by the sample standard deviation $\sigma$: $Z=(X-\mu)/\sigma$.}, the $\beta_{i,j}$ coefficients for different stars were taking similar values for different stars. This allows to define a model $Z_j(\lambda)$ of the $j^{\rm th}$ instrumental systematic, which can be applied to any star, without necessarily performing the PCA on that specific star's CBC RVs. 

For HARPS, we decided to do this for the two clear instrumental systematics identified in Sect.~\ref{sec:folding}. For each of them, we constructed a model by merging all the $Z-$scores of HARPS targets related to the component under consideration into a single array, then computing the median within a sliding wavelength window. The master calibration curves, $Z_1(\lambda)$ and $Z_2(\lambda)$ respectively, were obtained using a window of the same width as the one used to generate the CBC RVs (i.e.\ 4 \ang{}), and are displayed in Fig.~\ref{FigSlice1}. Note that, for $Z_2$, the pixel position could have been add to the model to fit a 2D smooth function $Z_2(\lambda,pixel)$, but here the effect is already quite clear with the wavelength $\lambda$ only. Once a function $Z_j(\lambda)$ is known, it can be used to form the groups on which the PCA will be fit as follows. 

For a specific star and a specific systematic $j$, we evaluate $Z_{i,j}$ at the location of each line $\lambda_i$ included in our tailored line selection by linearly interpolating the $Z_j$ master curve. The $Z_{i,j}$ values are taken as an estimate of the $\beta_{i,j}$ coefficients, and used to form 10 groups of equivalent size (cut at every $10^{th}$ percentile). The LBL RVs are averaged inside each group using Eq.\ref{eq:1}. This process allows us to go from a cloud of thousands of points at low S/N to a cloud of ten points at high S/N. We then perform PCA on this reduced cloud and extract the first principal component. Only the first component is relevant, since groups are precisely formed in a way that the variance is magnified along the variance of the expected systematic and averaged in the other direction. This process is equivalent to the one illustrated in Fig.\ref{FigPCA2}, where the color gradient is now given by the $Z_{i,j}$ values. We confirmed that this procedure produces a single significant PC by checking the explained variance curve of the PCs. 

Note that, an interesting property of the present analysis is that it does not require to observe the same sample of stars for all the epochs to correct for the underlying instrumental systematic. 

\subsection{Final refined corrections defining the end point of YARARA V2
\label{sec:iterations}}

In a previous paper \citep{Cretignier(2021)}, we developed a post-processing pipeline called YARARA dedicated to the flux corrections of known systematics at the spectrum level. From this improved version of the spectra, more precise LBL RVs were extracted and corrected by the shell decomposition presented in \citet{Cretignier(2022)}. The present paper comes as a further stage of corrections applied after the shell decorrelation and also applied on the LBL RVs. Since YARARA was dedicated to flux corrections, whereas shell and LBL PCA are time-domain corrections, we will call hereafter the final LBL RVs obtained after PCA correction the "YARARA V2" products (or YV2), as opposed to the RVs obtained after the flux correction in \citet{Cretignier(2021)}, which we call refer to as YARARA V1 (or YV1) RVs. 

After extracting the YV1 LBL RVs and performing the shell decomposition, the YV2 correction of the residual LBL RVs is performed in three consecutive stages:
\begin{enumerate}
    \item correct common-mode instrumental systematics using the master $Z_j(\lambda)$ calibration curves, as described in Sect.~\ref{sec:systematics}; 
    \item use the residual LBL RVs obtained after step 1.\ to construct CBC RVs, apply PCA decomposition to the latter, and use the resulting PCs to correct the LBL RVs
    \item apply PCA directly on the residual LBL RVs obtained after step 2.\ to perform a final correction for any effects that do not display a smooth wavelength dependence.
\end{enumerate}
The number of PCs fit in steps 2.\ and 3.\ was determined as in \citet{Cretignier(2022)}, using a leave-$p$-out cross-validation algorithm. However, we slightly modified the algorithm since such method was unstable numerically. The new method is described in Appendix \ref{appendix:a} and is closer to the version also used in \citet{Ould(2023)}. 
For convenience, from here onward we will refer to the PCs identified at steps 1., 2.\ and 3.\ as \textit{"slice"}, \textit{"color"}, and \textit{"lbl"}, respectively.

\subsection{Including Keplerians in the model
\label{sec:keplerians}}

As discussed in Sect.~\ref{sec:PCA}, the presence of a planetary signal has no impact on the PCA itself in that it does not affect the extracted PCs. However, the PCA correction can remove part of a planetary signal if the latter displays some linear correlations with one or more of the PCs. Naturally, this could hinder the detection of the planetary signal in question and affect the estimate of its parameters. Furthermore, as both the planetary model and the coefficients of the PCs ($\alpha_{i,j}$) are affected, the resulting combined model is imperfect. Fitting an imperfect model to the data can result in the injection of additional, spurious signals with small amplitudes that could be misinterpreted. 

An example of this effect arose during the planetary injection-recovery tests we performed on HD10700 (see Sect.~\ref{sec:hd10700}). One of the planets we injected, with semi-amplitude $K=3$\,\ms{} and period $P=122$ days (one third of a year), was interacting with the PCs containing power at 1-year periodicity. Consequently, the planetary amplitude was reduced by 40\% and a significant 1-year signal was visible after YV2, which was not present in the YV1 RVs. Such behaviour is clearly undesirable but can be avoided by fitting a Keplerian model simultaneously with the PCs. 

So far we neglected the planetary signal, $\text{RV}_P(t)$ in Eq.\ref{eq:11}, when evaluating the $\alpha_{i,j}$ coefficients. We now re-introduce it explicitly, as follows. For the purposes of estimating the $\alpha_{i,j}$'s, the precise functional form of the $\text{RV}_P(t)$ signal (which, in general, consists of the combined signal of several exoplanets) is not relevant: we are only interested in approximating it well enough in the time-domain. For that reason, 
we choose to fit a superposition of $C_k(t)$ circular orbits, which is more stable numerically and enables us to preserve the linearity of the model:
\begin{equation}
\label{eq:6}
\text{RV}_P(t) = \sum^n_{k=1} C_{k}(t) \equiv \sum^n_{k=1} A_{k} \cdot \sin\left(\frac{2\pi}{P_k} \cdot t\right) + B_{k} \cdot \cos\left(\frac{2\pi}{P_k} \cdot t\right)
\end{equation}
In this framework, the signals of planets with significant eccentricity would be captured by several components at the orbital period and its harmonics.
If the phase of the planetary signal is known (for example for transiting planets), the two terms in Eq.~\ref{eq:6} can be replaced by a single sinusoidal function with the appropriate phase. 

In principle, the coefficients $A_k$ and $B_k$ should be the same for all the stellar lines. However, the model 
was easier to implement if the coefficients were free to change from line to line, as the fit then proceeds on a line-by-line basis: 
\begin{equation}
\label{eq:10}
\text{RV}_i(t) = \sum^n_{k=1} A_{i,k} \cdot \sin\left(\frac{2\pi}{P_k} \cdot t\right) + B_{i,k} \cdot \cos\left(\frac{2\pi}{P_k} \cdot t\right) +  \sum^N_{j=1} \alpha_{i,j}\cdot \text{PC}_{j} (t)
\end{equation}
While this might not be as optimal as fitting a global set of $A_k$'s and $B_k$'s, we noted that the values of the $A_{i,k}$ and $B_{i,k}$ coefficients are not used to estimate the planet parameters. Their purpose is only to minimize crosstalk between any signal with a periodicity $P_k$ and the $\alpha_{i,j}$ fit coefficients. The final planet parameters are obtained by fitting Keplerian orbits to the YV2-corrected RVs after averaging the latter over the individual lines, as described below.

The above procedure assumes that the period(s) of the planetary signal(s) are known. This naturally begs the question of how to identify these signals, at the same time as fitting for the instrumental systematics. This is challenging not only because of the "red noise" imparted by the systematics, but also because several Keplerian signals can mix together due to the time sampling of the observations. This issue can be addressed by searching for signals at all periods \textit{simultaneously} using the l1-periodogram introduced in the context of RVs by \citet{Hara(2017)}. 

A convenient feature of the l1-periodogram is that, in addition to the periodic signals one is searching for, a basis of linear predictors can be included explicitly into the model. While the unknown periodic terms are penalized using L1 regularization to avoid over-fitting, the regularisation is not applied to the known basis terms.
 We therefore apply the l1-periodogram to the YV1 RV time-series, adding our PCs to the basis of \textit{unpenalized vectors}. We then keep all the periods $P_k$ with a False Alarm Probability (FAP) lower than 0.1\%, where the FAP is defined as in \citet{Delisle(2020)}. 

One important consideration when using the l1-periodogram is the noise model, which has a significant impact on the final power landscape. In the l1-periodogram, this is implemented via a user-specified covariance matrix. In this work, we used a purely diagonal covariance matrix, i.e.\ we assumed that the noise was white. The main expected noise contributions are photon noise, which is white, and (super-)granulation, which can be reasonably approximated as white on the timescales of our nightly-binned data. However, while the magnitude of the photon noise can be estimated, it can be more difficult to assess the (super-)granulation contribution, which depends on the observational strategy \citep{Meunier(2017)}. We estimate the overall white noise level empirically, by fitting an iterative Generalised Least-squared (GLS) periodogram \citep{Zechmeister(2009)}, using the same FAP level\footnote{Note that this time, the FAP is analytically computed as described from \citet{Baluev(2008)} which is strictly equivalent to \citet{Delisle(2020)} in the case white noise.} criterion of 0.1\%. At each iteration, we identify the highest peak in power, and add a sinusoid at that period to the model, using the Keplerian fitting code published in \citet{Delisle(2016)}. The iteration stops when the highest peak is above the specified FAP threshold. 
We then estimated the median absolute deviation of the RV residuals, and added this in quadrature to the theoretical (photon-noise) uncertainties for eac observation in our l1-periodogram analysis.

In practice, the set of periods detected in the l1-periodogram following the above prescription is not necessarily the optimal set to include in the final model fit. First, it can sometimes still include spurious periodicities induced by red noise components that were not filtered out by our vector basis (see for example the 15-day signal found in the RVs of HD109200, discussed in Sect.~\ref{sec:hd109200}). Furthermore, the l1-periodogram is known to struggle to detect low-amplitude signals in the presence of much larger ones \citep{Hara(2017)}. On the other hand, the iterative GLS procedure described above to estimate the excess white noise level also yields an alternative set of periodicities, which we found to be less affected by these two limitations.

Since the GLS implementation we used in this work does not allow us to fit a vector basis simultaneously with a sinusoid at each trial period, we first used the set of periods $P_k$ identified with the l1-periodogram to evaluate a preliminary YV2 systematics correction, and ran the iterative GLS on the residuals. As with the l1-periodogram analysis, we fit only for circular orbits, eccentric orbits therefore give rise to multiple detected periodicities at harmonics of each other. In this way, we obtain a new set of periods $P_k$, which may differ from the set identified with the l1-periodogram, though the largest signals are usually common to both. 

Our final model for the global RV time-series is then: 
\begin{equation}
\label{eq:8}
\text{RV}(t) = \sum^n_{k=1} C_{k}(t) + \sum^l_{i=1} \omega_i \sum^N_{j=1} \alpha_{i,j}\cdot \text{PC}_{j} (t) + \text{RV}_{res}(t)
\end{equation}
where the $C_k(t)$ are the periodic signals identified using the iterative GLS procedure, and the final $\alpha_{i,j}$ are re-estimated using this final set of periods.

We perform a further test on each of the candidate planetary signals $C_k(t)$ to determine which of them can be considered as a significant planet candidate and thus should be included in the final set of periods used in Eq.~\ref{eq:8}. This tests consists in comparing the candidate signal $C_k(t)$ with the residual signal $\text{RV}_{res}(t)$. The test works as follows: for each of the $k$ candidate signals, we first construct a systematics-corrected RV time-series $\text{RV}_k(t) = C_k(t)+\text{RV}_{res}(t)$, which includes only the planetary signal in question and the residuals of the full model. We then test whether the highest peak in the GLS periodogram of $\text{RV}_k(t)$ is still at period $P_k$ when a small fraction of the observations is deleted. If removing a small fraction of the observations changes the location of the highest periodogram peak, that peak was most likely due to residual correlated noise. On the other hand, (provided that the phase coverage is not strongly heterogeneous), a true planetary signal should be present in the whole dataset, and therefore removing a small fraction of the observations might reduce the significance of the highest peak, but should not change its period altogether. This approach is conceptually similar to the apodised periodogram of \citet{Hara(2022)}, which tests how the strength of a signal changes over time. 

To implement this test, we had to decide what fraction of the data points to remove and how to select them. We arbitrarily set the fraction to 10\% of the total. Most of the datasets considered in the present study consist of $\sim\,300$ observations, so that roughly $\sim\,30$ observations were rejected. We decided to remove the observations with the largest values of $|C_k(t) \cdot \text{RV}_{res}(t)|$, i.e.\ those where the candidate planetary signal was the largest (closest to quadrature). This allows us to verify that the signal is not produced by outliers falling around quadrature. We also investigated other ways of selecting the points to remove, including selecting points at random or according to S/N ratio, but ultimately opted for the "\textit{quadrature attack}" as it is the most aggressive, and therefore the set of candidate signals that pass this set should be the most robust. Note that we did not use the value of the peak power or FAP in this quadrature attack process, we only test whether the peaks remain the highest, i.e.\ whether the period that best explains the candidate signal is robust to the removal of the selected points.

Once all the periods $P_k$ were tested, only those surviving \textit{quadrature attack} test were kept. We then re-evaluated the $\alpha_{i,j}$ coefficients following Eq.\ref{eq:10}, and used them to correct the $\text{RV}_i(t)$ before using Eq.\ref{eq:1} to compute the weighted average of the corrected LBL RVs. The resulting $\text{RV}(t)$ is the final YV2 time-series presented for each of our targets in Sect.~\ref{sec:results}. All the time-series (YV1, YV2 and PCs) are provided in machine readable format in the Supplementary Online material\footnote{Give the link to the extra material}. 

The procedure described in this subsection to identify and refine the set of planetary signals to include in the final fit may seem complex and somewhat ad-hoc, but it reflects the challenge of finding an automatic routine that can be applied on large database. Ultimately, a human still needs to check and carefully investigate each of the periods included in the final set. 

\begin{figure*}[t]
	
	\centering
	\includegraphics[width=16.8cm]{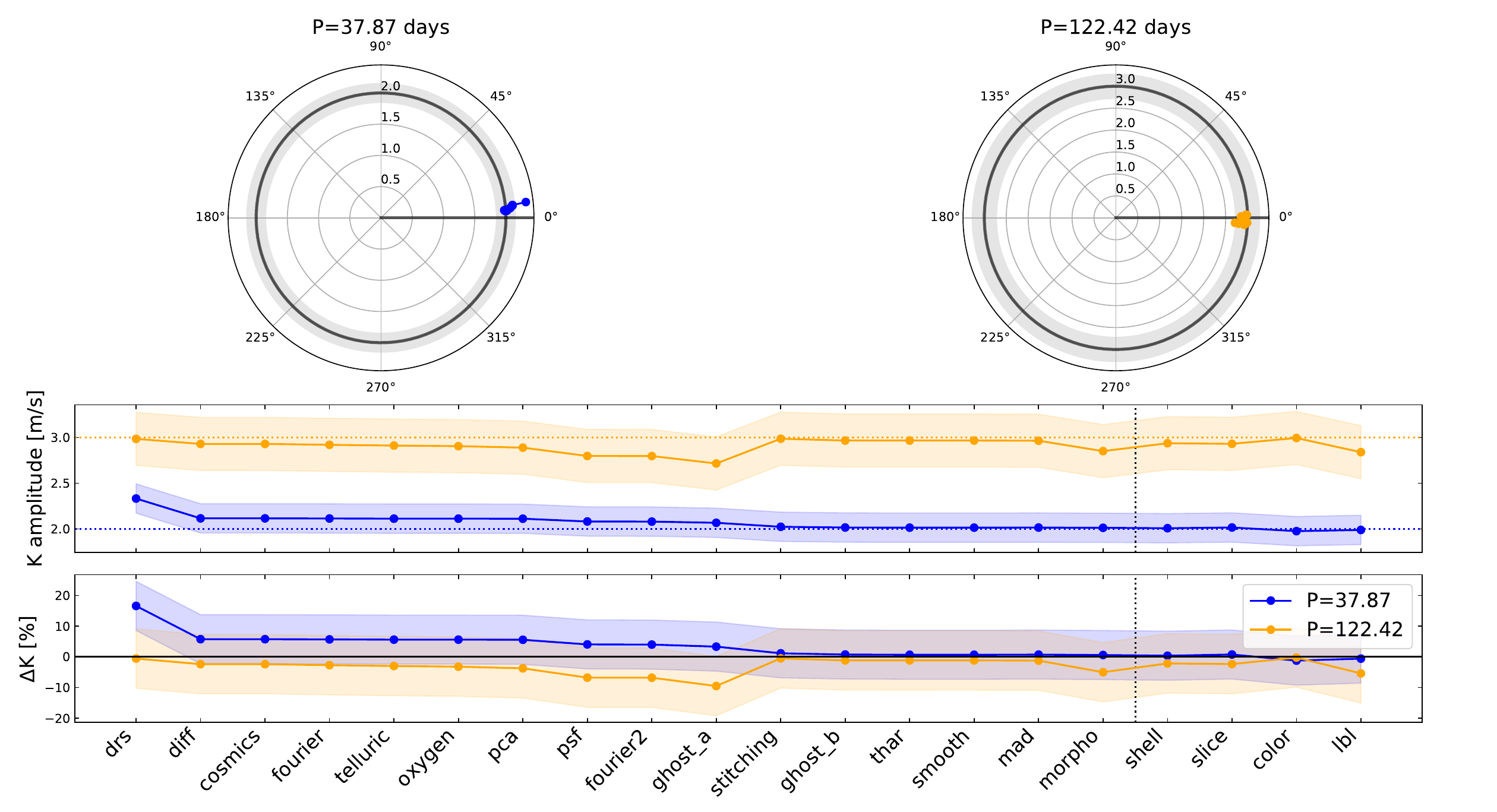}
	\caption{
	Planetary absorption along the YARARA V1 and V2 sequence of the planetary injection test performed on HD10700. Planets at 37 and 122 days were injected at the spectrum level. A 2-Keplerian circular model at the periods of the injected planet was fit at each step of the pipeline to trace the recipes with potential absorption facilities. An uncertainty of 10\% on the $K$ semi-amplitude was typical expected due to the presence of smaller amplitude planetary candidates \citep{Feng(2017),Cretignier(2021)}. \textbf{Top:} Representation of the 2-circular solution fit into a polar diagram, the radius represents the $K$ semi-amplitude and angle the phase of the signal. Injected amplitudes (2 and 3\,\ms{}) are highlighted by the dark circle, and reference phase were at null value. \textbf{Middle:} Variation of the fit K-semi amplitudes as a function of the YARARA's stage. Separation between YV1 and YV2 is depicted (vertical black dotted line). The shell correction \citep{Cretignier(2022)} is considered as the first stage of YARARA V2, whereas the three steps corrections described in Sect.~\ref{sec:iterations} are labeled as \emph{slice}, \emph{color} and \emph{lbl} respectively. \textbf{Bottom:} Same as middle in relative percentage difference. At the end of YV2, the recovered amplitudes and phases are plainly compatible with the injected ones.}
	\label{FigInjection}
	
\end{figure*} 

\section{Results
\label{sec:results}}

We present the results of the extended version of YARARA obtained on five intensively observed HARPS targets: HD10700, HD192310, HD115617, HD109200 and HD20794. 
All the orbital solutions were obtained using the publicly available code\footnote{\url{https://pypi.org/project/kepmodel/}} to fit Keplerians developed by \citet{Delisle(2016)}.

We used HD10700 as a planetary injection test since this star does not contains any large RV signal from planets or stellar activity and is a standard test to validate the capability of a method to not absorb planetary signals. The next star, HD192310, was chosen since it contains two clear exoplanets and a third peak related to the rotational period at 44 days. 

The third star, HD115617, present an interesting case of planetary signals in interaction with systematics. Moreover, the present work tends to show that the mass of the third published exoplanets was overestimated by a factor two.
The fourth star, HD109200, is a complex case of stellar activity signal unresolved with the present work. It is also the lowest S/N dataset, which shows the difficulty to work with signal-to-noise ratio lower than 200.
Finally, the last star, HD20794, was selected since it contains also two clear planetary signals whereas a third one seems to be a valid Super-Earth planetary candidate. The current MCMC solution converges to an eccentric orbit crossing the habitable zone of this solar-type star (G6V).  

\subsection{Planetary injection on HD10700
\label{sec:hd10700}}

We tested our YARARA V2 pipeline on the same planetary injection dataset than in \citet{Cretignier(2021)}. As a reminder, HD10700 is one of the star intensively observed by HARPS presenting the lowest RV rms on the full lifetime of the instrument ($\sim$1\,\ms{}). The star was observed during 380 nights between $1^{st}$ August 2005 and $18^{th}$ December 2014. This target is of primary interest to test our method since no large signal coming either from planets, or from the stellar activity is observed. Two circular-orbits planets were injected directly at the spectrum level (see \citet{Cretignier(2021)}), where one of the planet ($P=37$ days and $K=2$\,\ms{}) was not expected to interact with any recipes of YARARA and is a sanity check, whereas the second one was expected to crosstalk with 1-year systematics ($P=122$ days and $K=3$\,\ms{}). 

We already demonstrated in \citet{Cretignier(2021)} that the amplitudes recovered at the end of YARARA V1 were compatible with the injected ones considering a 7\% uncertainty due to the potential presence of lower planetary signals \citep{Feng(2017)}, validating the ability of the recipes performed at the flux level to not absorb the planetary signals. We propose now to test the ability of YARARA V2 to perform a similar job, which is more challenging to achieve in the time-domain. In total, YV2 includes 13 vectors fit in the time-domain (4 \textit{shells}, 2 \textit{slices}, 3 \textit{colors} and 5 \textit{lbl} according to the nomenclature defined in Sect.~\ref{sec:iterations}). This large amount of vector is due to the exquisite S/N of the observations ($med(S/N)=512$). Such a number of components could look like a large basis, but given that 380 epochs are used, a multi-linear model has in fact a low risk of over-fitting the data. 

When running the first iteration of YARARA V2, we found out that the amplitude of the 122-days planet was strongly absorbed. The 4 shells vectors first decrease its amplitude to 2.75\,\ms{}, whereas the signal was decreased down to 1.50\,\ms{} once the full basis of 13 vectors were fit. Both planets were still detectable in a GLS periodogram but a third significant peak at 1-year not present in YARARA V1 was now visible with a semi-amplitude $K$ around 1\,\ms{}. It turned out that several PCs components correcting some of the 1-year systematics of the instrument were attempting to absorb the 122-days planet. This is an expected behaviour from cross-term between planetary signals and PCs basis since the $\text{RV}_P(t)$ model in Eq.\ref{eq:3} was set to zero at the beginning. 

This example shows that GLS should be carefully interpreted when applied on a residual RV time-series filtered out by some red noise model. Despite the high collinearity, the 122-day signal was perfectly identified in the l1-periodogram, as well as a 20-day and the 37-day signals. Hence, the periods were therefore added in the model (Eq.\ref{eq:10}). The signals were now no more absorbed and both fulfilled the quadrature attack test described in Sect. ~\ref{sec:keplerians} which preserved them in the Keplerian model.

\begin{figure*}
	
	\centering
	\includegraphics[width=17.4cm]{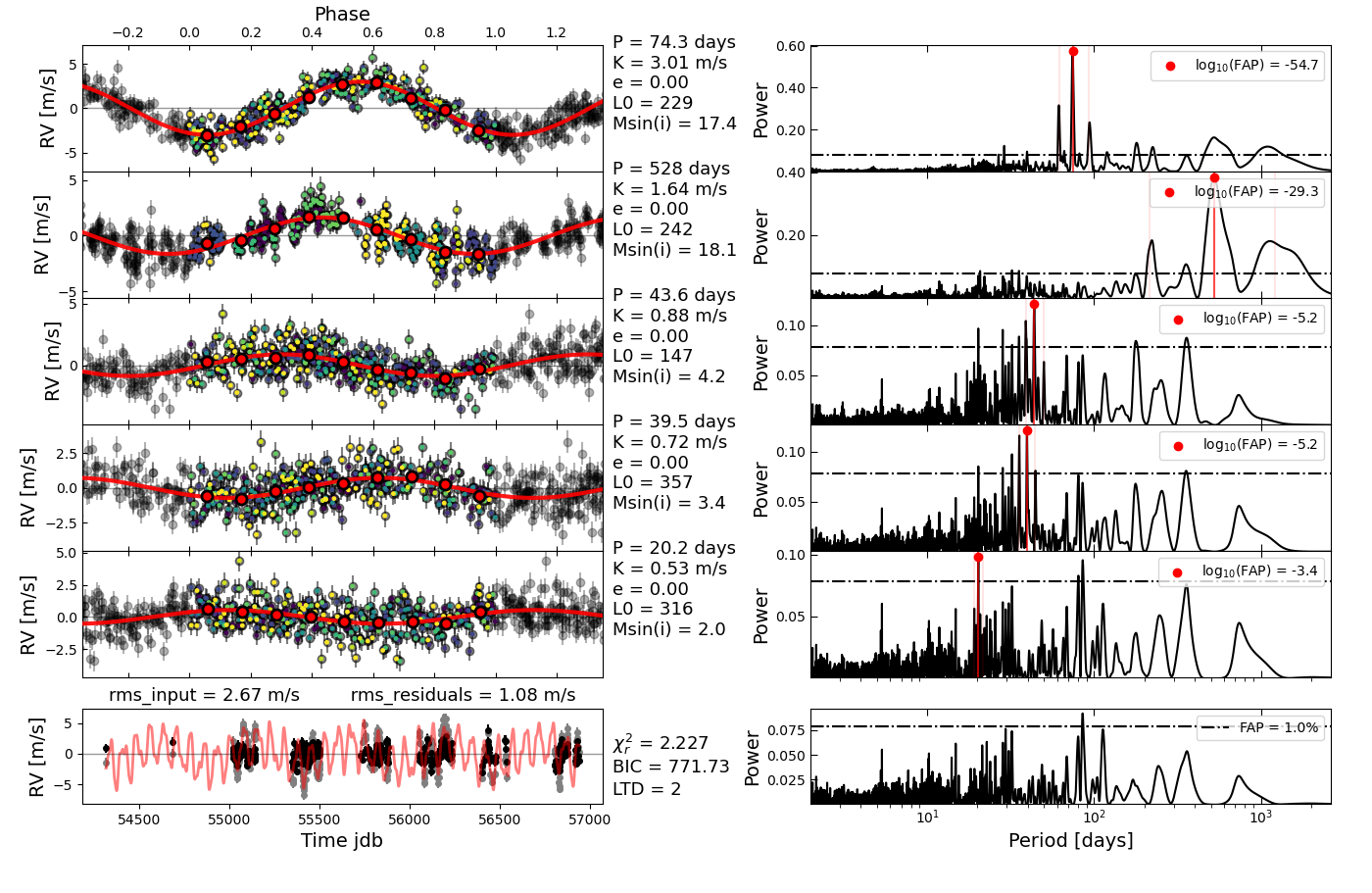}
	\caption{
	Iterative circular orbits Keplerian-fit (from top to bottom) on HD192310 for the RV time-series obtained with the DRS. The final Keplerian solution around the residuals is showed on the left. Model (red curve), data (grey dots) and residuals (black dots) are displayed at the bottom.}
	\label{FigKep3}
    
	\centering
	\includegraphics[width=17.4cm]{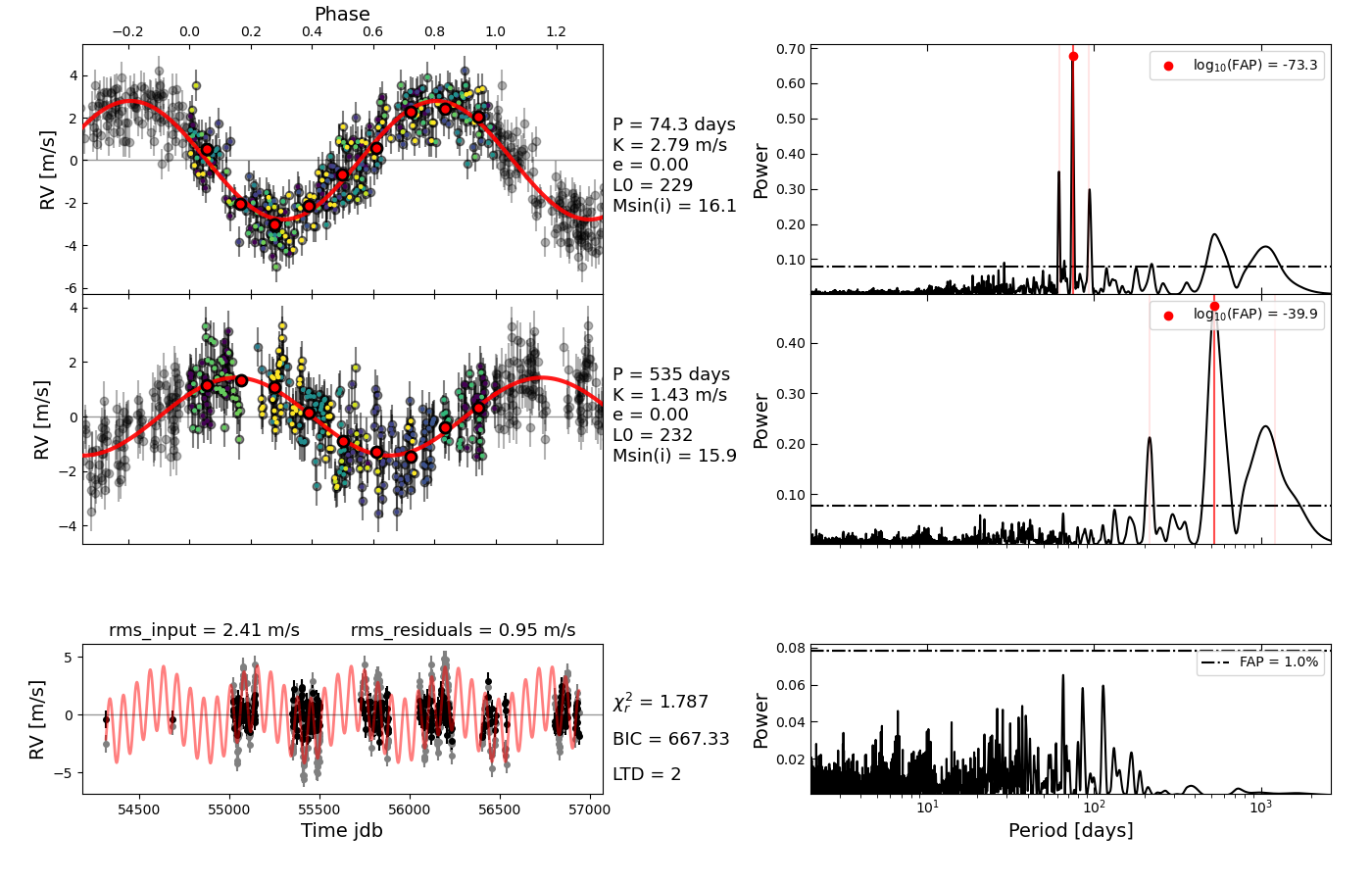}
	\caption{Same as Fig.~\ref{FigKep3} for the YARARA V2 RV time-series. No significant extra signals around 44 days is detected. The residuals around the model is now of 94\,\cms{}. We kept the quadratic long-term drift (LTD)  for fair model comparison.}
	\label{FigKep4}
	
\end{figure*}

At the end of YARARA V2, both planetary signals were perfectly recovered with the right phases and right amplitudes with values of 1.99 and 2.89\,\ms{} respectively, equivalent to a relative difference of 1 and 5\% with the injected ones. We fit at each stage of the YV1 and YV2 sequence, a two-Keplerian model of circular orbits with the known periods of the injected signals in order to follow and evaluate the evolution of the phases and amplitudes of the signals during the sequence. This analysis is represented in Fig.~\ref{FigInjection}. We observed that for any recipe in YV1 and YV2, the amplitudes (or the phases) deviates farther away than 10\% of the injected signals' relative values.

We therefore conclude that the present version of the pipeline is able to recover planetary signal without significant absorption of their amplitudes neither crosstalk power communication. A single exception must be noted concerning long term signals, which are by nature highly collinear and degenerate with any other long trend signal. The method presented here therefore assumes that the baseline of the observation covers, as a rule of thumb, at least twice the planetary period, which is in general a quite common requirement to the publication of any signal with an amplitude close to the instrumental limit.

\subsection{HD192310}
\label{sec:hd192310}

HD192310 is a bright K2V star ($m_v=5.7$) located at 8.8 pc of the Sun with a stellar mass of 0.82 \Msun{} \citep{Ramirez(2013)}. With such a high stellar magnitude, the median signal-to-noise ratio of the observations at 5500 \ang{} is good and raises to $med(S/N)=307$. The star exhibits a magnetic cycle with a periodicity estimated around $\sim9.5$ years based on the CaII index. Moreover, a rotational period between 39 and 45 days can be estimated from several classical activity proxies as CaII H\&K lines, $H_\alpha$ or the CCF VSPAN, values compatible with the reported rotational period found in \citet{Pepe(2011b)}.

The star has been intensively observed as part of the HARPS large program with 322 nightly observations between 4 August 2007 and 6 October 2014. The RVs revealed two convincing Neptune-like exoplanets at 75 and 525 days \citep{Pepe(2011b)} with semi-amplitudes larger than 2\,\ms{}. A 60\,\cms{} signal at 25 days was also published as a candidate exoplanet recently by \citet{Laliotis(2023)}.

By performing an iterative Keplerian fit on the RV time-series coming from the DRS (Fig.~\ref{FigKep3}), we detected the same clear signatures around 74 and 528 days with amplitudes of 3.02 and 1.64\,\ms{} respectively. Note that an extra parabolic trend component was needed to fit the data due to the magnetic cycle. 
After subtraction of these two signals, three other signals are detected at 44, 39 and 20 days with amplitudes around $\sim$0.80\,\ms{}.  
Except the clear peak visible at 44 days in several chromospheric activity proxies and CCF moments (once the magnetic cycle removed), the application of YARARA further confirms the parasite nature of those signals and show an origin certainly driven by the stellar activity. Note that other signals can be detected but, we stopped the iterative fit at 5 planets. Such a large number of planets detected at the instrumental precision level is symptomatic of red noise components that are imperfectly fit by sinusoidal functions and that need several Keplerian to be approximated by the model.

After the flux correction of YV1, a total of 11 vectors were fit in the time-domain (7 \textit{shell}, 2 \textit{slice}, 1 \textit{color} and 1 \textit{lbl} components). On the YV2 dataset, the planetary signal at 75 and 550 days are recovered as displayed in Fig.~\ref{FigKep4} and the stellar activity signals around the rotational period at 44 days are now no more significant, a result similar to the one found in \citet{Cretignier(2022)} for the K1V dwarf star $\alpha$ Cen B. The same semi-amplitude of 2.79\,\ms{} and 1.43\,\ms{} are obtained for the 75 days and 550 days signal respectively, demonstrating once again (on real planetary signals now) that PCA performed on LBL RVs and CBC RVs are indeed insensitive to planetary signals and planetary absorption only occurs due to partial linear correlation with the basis fit. The periodogram of the residuals is already much cleaner and only present peaks at 95 days and aliases below the 1\% FAP level.

In the residuals of YARARA V2, we did not find a hint of the planetary candidate mentioned in \citet{Laliotis(2023)} at 24 days. In that study, the authors do not properly model out the stellar activity and simply fit it out via a Keplerian solution. Fitting a non-Keplerian signal such as stellar activity or instrumental systematics with a Keplerian signal can easily introduce other peak in periodograms, which may explain this extra signal, in particular since the semi-amplitude of the candidate is around 60\,\cms{}. Furthermore, all the HARPS systematics are still present in their data \emph{a priori} and the peak-to-peak for some of them are easily above 60\,\cms{} as shown in \citet{Cretignier(2021)}. Their analysis also includes the RV time-series of other instruments such as UCLES and HIRES, but none of them have the sensitivity to detect such a small signal and including them with HARPS dataset would have increased the red noise model complexity (since each new instrument brings its own pack of systematics) rather than improving the RV precision. 

\renewcommand{\arraystretch}{2.0}
\begin{table*}[tp]
\caption{Orbital and physical parameters obtained from the MCMC performed on the different RV time-series of HD192310 for a 2-Keplerian fit model. The reference date is $BJD=2'455'500$ and the stellar mass is taken as $M_*=0.82$ \Msun{} \citep{Ramirez(2013)}. Units are in days for the period $P$,\,\ms{} for the semi-amplitude $K$, degree for the periastron angle $\omega$ and for the node angle $\lambda_0$, AU for the semi-major axis $a$ and \Mearth{} for the minimum mass $m$ $\sini{}$.}
\label{Table}
\centering
\begin{tabular}{c|cc|cc|cc}
\hline\hline
 & \multicolumn{2}{c|}{DRS} & \multicolumn{2}{c|}{YV1} & \multicolumn{2}{c}{YV2} \\
\hline
Par. & planet b & planet c & planet b & planet c & planet b & planet c \\
\hline
$P$ & $74.16^{+0.04}_{-0.04} $ & $528.3^{+3.6}_{-3.8} $ & $74.27^{+0.04}_{-0.04} $ & $528.4^{+5.1}_{-5.2} $ & $74.25^{+0.04}_{-0.04} $ & $534.9^{+5.9}_{-5.1} $ \\
$K$ & $2.94^{+0.06}_{-0.06} $ & $1.78^{+0.08}_{-0.08} $ & $2.93^{+0.06}_{-0.06} $ & $1.51^{+0.07}_{-0.07} $ & $2.81^{+0.06}_{-0.06} $ & $1.44^{+0.06}_{-0.06} $ \\
$e$ & $0.11^{+0.02}_{-0.02} $ & $0.31^{+0.05}_{-0.05} $ & $0.13^{+0.02}_{-0.02} $ & $0.17^{+0.05}_{-0.05} $ & $0.11^{+0.02}_{-0.02} $ & $0.06^{+0.05}_{-0.04} $ \\
$\omega$ & $161^{+11}_{-13} $ & $14^{+8}_{-7} $ & $157^{+9}_{-10} $ & $78^{+15}_{-15} $ & $151^{+12}_{-13} $ & $123^{+51}_{-44} $ \\
$\lambda_0$ & $229^{+1}_{-1} $ & $241^{+3}_{-3} $ & $230^{+1}_{-1} $ & $243^{+4}_{-4} $ & $229^{+1}_{-1} $ & $235^{+4}_{-4} $ \\
\hline
$a$ & $0.32^{+0.01}_{-0.01} $ & $1.20^{+0.02}_{-0.02} $ & $0.32^{+0.01}_{-0.01} $ & $1.20^{+0.02}_{-0.02} $ & $0.32^{+0.01}_{-0.01} $ & $1.21^{+0.02}_{-0.02} $ \\
$m$ \sini{} & $16.8^{+0.7}_{-0.7} $ & $18.8^{+1.0}_{-1.0} $ & $16.8^{+0.7}_{-0.7} $ & $16.5^{+0.9}_{-0.9} $ & $16.1^{+0.7}_{-0.7} $ & $15.9^{+0.9}_{-0.9} $ \\
\hline
\end{tabular}
\end{table*}

We ran a Monte-Carlo Markov chain (MCMC) to fit a 2-Keplerian model with flat priors using the available algorithm on the DACE platform\footnote{\url{https://dace.unige.ch}}. The MCMC implementation is based on \citet{Diaz(2014),Diaz(2016)}. The chain was made of 860'000 iterations and the first quarter was burnt. The resulting table for the posterior distribution of the orbital parameters is displayed in Table.\ref{Table}. Using the YV2 Keplerian solution, the RV rms of the residuals was decreased from 1.34\,\ms{} with the DRS dataset down to 0.90\,\ms{} for YV2.

Interestingly enough, we can note that the eccentricity of the planet HD192310 c is strongly reduced down to $e=0.11^{+0.02}_{-0.02}$, compared to $e=0.33\pm0.11$ in \citet{Pepe(2011b)}. Such a decrease of large eccentricity exoplanet was also observed for HD10180 f in our previous paper \citep{Cretignier(2021)}, a planet for which the orbital parameters are very similar to HD192310 c. Furthermore, the amplitude recovered is $K=1.44$\,\ms{}, which is also 37\% smaller than the one previously reported in \citet{Pepe(2011a)} at $K=2.27$\,\ms{}. We observe that the uncertainty on the parameters does not improve. This is due to the fact that for all the datasets, the same noise model is used (70\,\cms{} of white noise). This example show that orbital parameters obtained with uncorrected red noise modelling can be \textit{biased}, which could be concerning for some published signals close to the instrumental stability with amplitudes $K<2$\,\ms{}.

\begin{figure}
	\centering
	\includegraphics[width=9cm]{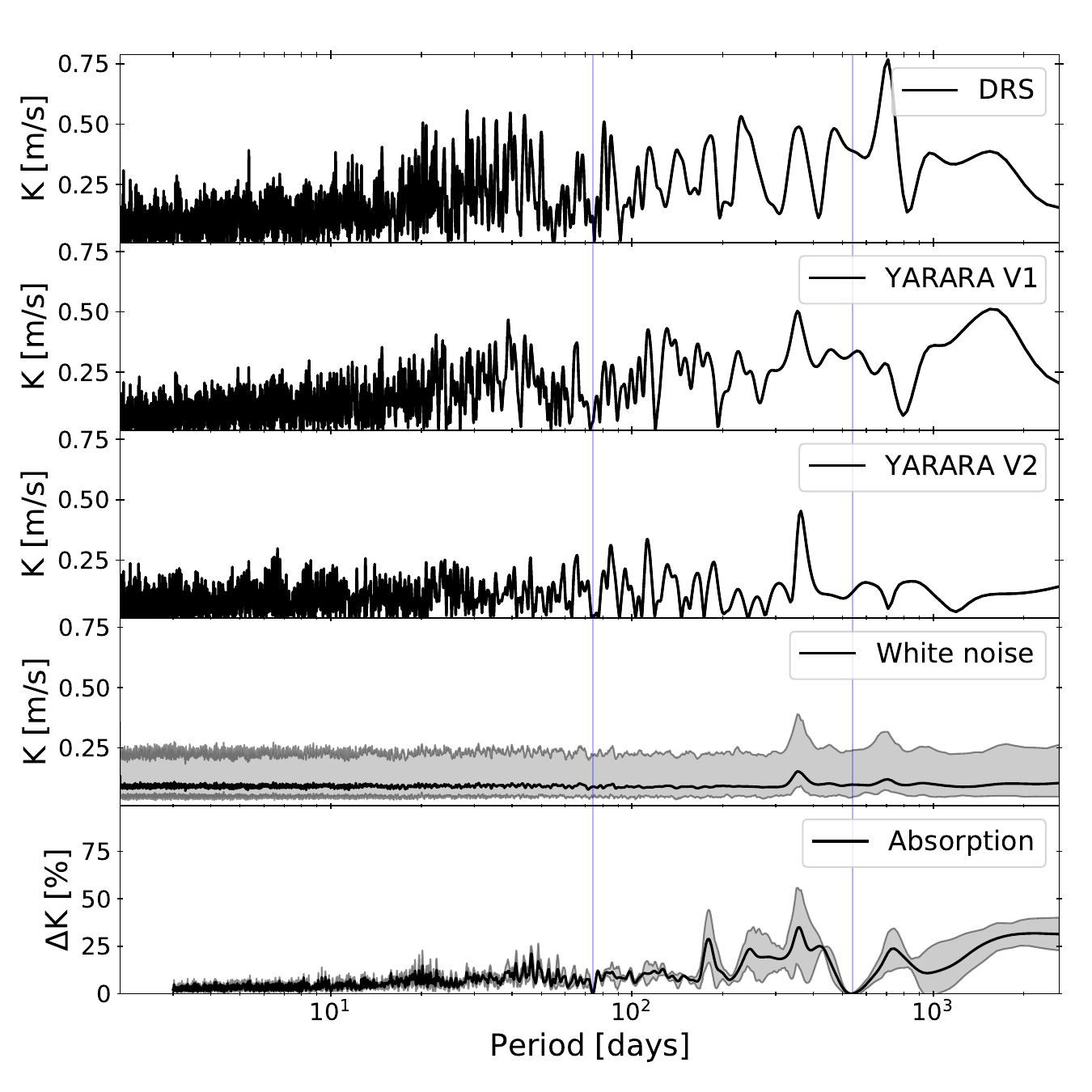}
	\caption{GLS periodogram of the different RV datasets for HD192310. Periodograms are plotted in $K$ semi-amplitude rather than power. The Keplerian solution obtained with the YARARA V2 dataset was removed. Position of the two planets is indicated by the vertical blue lines. \textbf{First row}: DRS RV time-series. Several $\sim$50 \cms contamination signals are detected. \textbf{Second row}: YV1 RV time-series. Power excess around 800 days, due to the stitching, was strongly mitigated. \textbf{Third row}: YV2 RV time-series. No strong residuals signal remains except a 45\,\cms{} 1-year signal already visible in the previous datasets. \textbf{Fourth row}: Distribution of 100 independent white noise realisations with a similar dispersion as the YV2 dataset. The mean (solid line) and 1-sigma ($16^{th}$ and $84^{th}$ percentiles as shaded area) are plotted. \textbf{Fifth row}: Absorption curve obtained by projecting sinusoidal curves on the vector basis. The solid line represents the mean absorption for the 18 phases tested, whereas the envelop show the $16^{th}$ and $84^{th}$ percentile.}
	\label{FigGlobalHD192310}
	
\end{figure}

In order to appreciate the improvement obtained with YARARA V2, we displayed in Fig.~\ref{FigGlobalHD192310}, the periodograms in amplitude of all the datasets, after removing the YV2 Keplerian solution of Table.\ref{Table}. Under the hypothesis that this solution is the good one, those periodograms allow to directly assess the contamination level in all the datasets. As an example, we observe a 65 \cms signal at 800 days strongly corrected already in YV1 and which mainly explain the large eccentricity of the HD192310 c with the DRS dataset. The improvement from YV1 to YV2 is also clear with less contamination around the rotational period.

In YV2, a peak at 1-year of 45\,\cms{} is visible, where such a level was already visible in the DRS and YV1 dataset as well. This peak can be induced by the window function of observational seasons, that tends to produce, even in presence of white noise, 1-year power because of the imperfect phase coverage. We showed that aspect by computing the periodogram for 100 independent white noise realisations with a dispersion identical to the YV2 dataset. In average, the amplitude at 1-year is around 15\,\cms{}, smaller than our 45\,\cms{} value, which indicate that some residual signals may have survived even after YV2, however such an amplitude of 40\,\cms{} rise up for 16\% ($84^{th}$ percentile) of the white noise realisations.

In order to demonstrate that the improvement from YV1 to YV2 is not at the expense of any absorption of real signals, we also displayed the \textit{absorption curve} as described in \citet{Cretignier(2022)}. As a reminder this curve is obtained by projecting on the PCs vector basis used to correct the RVs, sinusoidal curves sampled at the time of the observations. The amplitude of the curves after the projection is compared with the initial one to obtain the amplitude absorption $\Delta K$ in percent. For the simulation, 10'000 periods equidistant between 3 days and the baseline of the observations were tested, as well as 18 equidistant phases for each period.

For signal, shorter than 180 days, no absorption larger than 10\% is monitored. The largest absorption is around 1-year with $\Delta K=35\pm20\%$ depending on the phase of the injected signal. It indicates that several vectors from the basis contain 1-year power due to instrumental systematics. We can also see 0\% absorption at the location of the planets, this is because the planetary periods $P_k$ were added in the model (Eq.\ref{eq:10}) and fit simultaneously as the $\alpha_{i,j}$ coefficients which prevents any absorption at those periods.   

We preferred to use, as a figure of merit, the periodogram in amplitude rather than the extremely common root-mean-square (RMS) of the RV time-series residual. Indeed, RMS does not contain any information about the baseline of the time-series, the sampling of the observations, nor information about the remaining red noise frequency. All this information is however contained in the periodogram in amplitude. Nevertheless, there are drawbacks for periods poorly covered in phase by the observations (usually at 1-year), that may present arbitrary large amplitude since unconstrained. The present analysis is qualitative and not quantitative. By eye, we can see that on average, no signal larger than 25 \cms subsists after YV2. We highlight that this value, does not prevent the existence of planetary candidate with amplitudes larger than this value since those signals can still be mixed with residual red noise components. 

\begin{figure*}
	
	\centering
	\includegraphics[width=17.5cm]{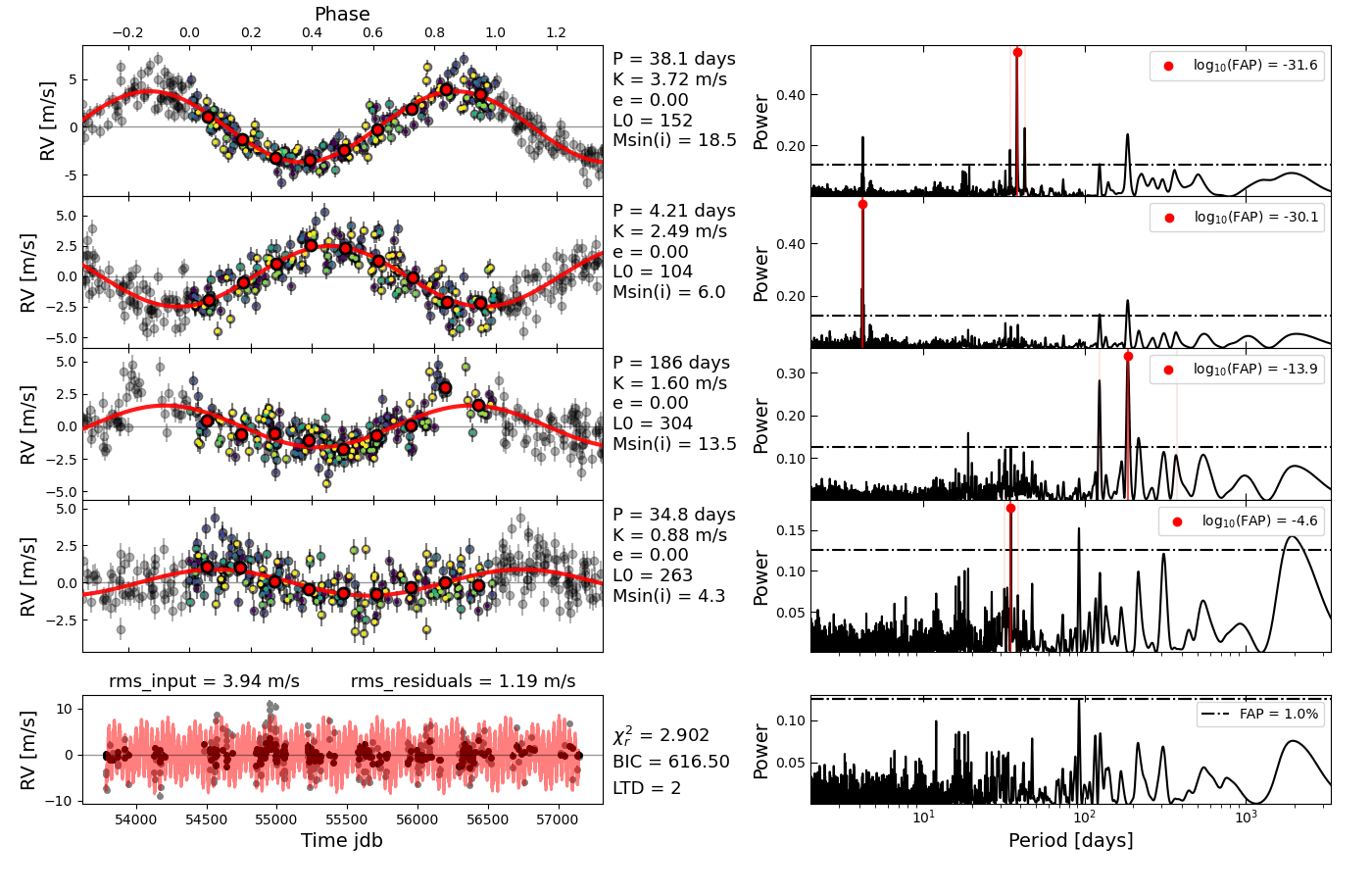}
	\caption{
	Iterative circular orbits fit of HD115617 with the RV time-series of the DRS. Exoplanets 61\,Vir, b and c a are detected but not 61\,Vir, d.}
	\label{Fig61Vir1}


	\centering
	\includegraphics[width=17.5cm]{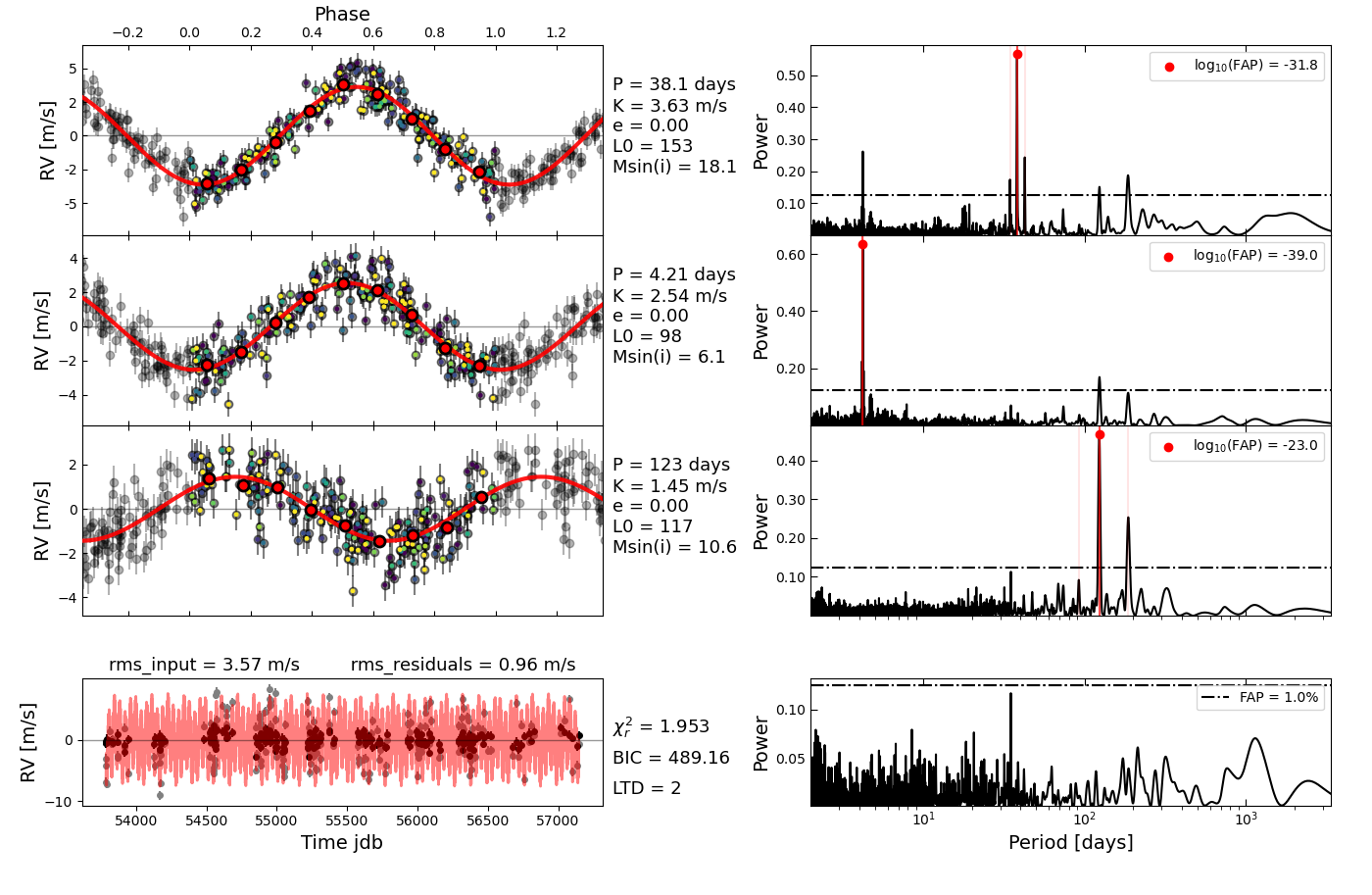}
	\caption{
	Same as Fig.~\ref{Fig61Vir1} with the RV time-series from YARARA V2. The planetary system of \citet{Vogt(2010)} is recovered.}
	\label{Fig61Vir2}
	
\end{figure*}

We outline three bullet points with the present example: 1) the planetary signals are well recovered and their significance are increased, 2) the rotational period around 44 days is now no more significant as for $\alpha$ Cen B in \citet{Cretignier(2022)} and 3) the residuals RV rms reached with the 2-Keplerian fit model is about 0.90\,\ms{} on 6 years of HARPS observation, significantly better compared to the 1.08\,\ms{} reached with the 5-Keplerian fit model obtained with the DRS and closer to the intrinsic limit of the instrument around 0.70\,\ms{}.

\subsection{HD115617}
\label{sec:hd115617}

HD115617, also called 61\,Vir, is a bright ($m_v=4.7$) G6V star located at 8.3 pc of the Sun with a stellar mass of 0.92 \Msun{} \citep{Ramirez(2013)}. The star has been observed 201 nights with HARPS between $19^{th}$ February 2006 and $5^{th}$ May 2015. The median S/N of the observations at 5500 \ang{} is $med(S/N)=387$. 

The star exhibits an irregular magnetic cycle based on the S-index of the CaIIH\&K lines with a periodicity close to 8 years. The star also showed a brutal outburst of its activity level in 2009 that may match with the maximum of the magnetic cycle. Once the long-term modulation filtered out, the periodogram of the S-index presents two clear peaks at 30 and 33 days likely related to the rotational period of the star. Three exoplanets were first detected and published by HIRES \citep{Vogt(2010)} with periods $P$ of 4, 38 and 124 days and $m_p \sin(i)$ projected mass of 5.1, 18.2, and 24.0 \Mearth{}.

\renewcommand{\arraystretch}{2.0}
\begin{table*}[tp]
\caption{Same as Table.\ref{Table} for HD115617 and for a 3-Keplerian fit model. The reference date is $BJD=2'455'500$ and the stellar mass is taken as $M_*=0.92$ \Msun{} \citep{Ramirez(2013)}. }
\label{Table3}
\centering
\begin{tabular}{c|ccc|ccc|ccc}
\hline\hline
 & & DRS & & & YV1 & & & YV2 & \\
\hline
Par. & planet b & planet c & planet d & planet b & planet c & planet d & planet b & planet c & planet d \\
\hline
$P$ & $4.215^{+0.001}_{-0.001} $ & $38.07^{+0.01}_{-0.01} $ & $123.5^{+0.2}_{-0.2} $ & $4.215^{+0.001}_{-0.001} $ & $38.09^{+0.01}_{-0.01} $ & $123.4^{+0.2}_{-0.2} $ & $4.215^{+0.001}_{-0.001} $ & $38.09^{+0.01}_{-0.01} $ & $123.2^{+0.2}_{-0.2} $ \\
$K$ & $2.60^{+0.08}_{-0.07} $ & $4.00^{+0.09}_{-0.09} $ & $1.38^{+0.25}_{-0.12} $ & $2.54^{+0.07}_{-0.07} $ & $3.47^{+0.07}_{-0.08} $ & $1.58^{+0.09}_{-0.09} $ & $2.53^{+0.07}_{-0.07} $ & $3.61^{+0.08}_{-0.07} $ & $1.44^{+0.08}_{-0.08} $ \\
$e$ & $0.03^{+0.03}_{-0.02} $ & $0.13^{+0.02}_{-0.02} $ & $0.17^{+0.31}_{-0.12} $ & $0.04^{+0.03}_{-0.03} $ & $0.04^{+0.02}_{-0.02} $ & $0.24^{+0.05}_{-0.06} $ & $0.05^{+0.03}_{-0.03} $ & $0.06^{+0.02}_{-0.02} $ & $0.12^{+0.05}_{-0.05} $ \\
$\omega$ & $161^{+76}_{-63} $ & $5^{+9}_{-8} $ & $324^{+13}_{-70} $ & $109^{+53}_{-42} $ & $339^{+31}_{-34} $ & $294^{+12}_{-12} $ & $80^{+34}_{-41} $ & $342^{+22}_{-22} $ & $253^{+29}_{-34} $ \\
$\lambda_0$ & $99^{+2}_{-2} $ & $150^{+1}_{-1} $ & $133^{+4}_{-4} $ & $101^{+2}_{-2} $ & $150^{+1}_{-1} $ & $116^{+3}_{-3} $ & $98^{+2}_{-2} $ & $152^{+1}_{-1} $ & $119^{+3}_{-3} $ \\
\hline
$a$ & $0.05^{+0.01}_{-0.01} $ & $0.22^{+0.01}_{-0.01} $ & $0.47^{+0.01}_{-0.01} $ & $0.05^{+0.01}_{-0.01} $ & $0.22^{+0.01}_{-0.01} $ & $0.47^{+0.01}_{-0.01} $ & $0.05^{+0.01}_{-0.01} $ & $0.22^{+0.01}_{-0.01} $ & $0.47^{+0.01}_{-0.01} $ \\
$m$ \sini{}  & $6.2^{+0.3}_{-0.3} $ & $19.7^{+0.8}_{-0.8} $ & $9.9^{+0.9}_{-0.8} $ & $6.1^{+0.3}_{-0.3} $ & $17.3^{+0.7}_{-0.7} $ & $11.3^{+0.8}_{-0.7} $ & $6.1^{+0.3}_{-0.3} $ & $17.9^{+0.8}_{-0.7} $ & $10.5^{+0.8}_{-0.7} $ \\
\hline
\end{tabular}
\end{table*}

61\,Vir, d is of a high interest since we can found some debate in literature around the 124-day signal. Indeed, initially, this candidate was not detected by HARPS \citep{Wyatt(2012)} and was even recently classified as a false positive \citep{Rosenthal(2021)}, before to be reconfirmed in \citet{Laliotis(2023)}. Another reason to investigate this signal is that planets close to 1-year harmonics may have biased orbits due to 1-year systematics present on HARPS, as showed in a previous paper \citep{Cretignier(2021)}. Such systematics are not exclusive to this spectrograph and are likely existing for other instruments as well. We further confirm such tendency with the present example in a more extreme case. Indeed, using only the HARPS data, iterative circular orbits converges towards a 180-day signal rather than the 124-day published planet (see Fig.~\ref{Fig61Vir1}). Quadratic trend was included in the model. Furthermore, a fourth signal at 35 days is also detected which is unlikely considering the Neptune-like planet 61\,Vir, b at 38 days. The RV rms of the 4-Keplerian fit is of 1.19\,\ms{}.

The planet at 120 days is already recovered in the YV1 RVs. After the flux correction of YARARA V1, a total of 11 vectors were fit in the time-domain (5 \textit{shell}, 2 \textit{slice}, 2 \textit{color}, 2 \textit{lbl}). We displayed in Fig.~\ref{Fig61Vir2}, the Keplerian iterative fit obtained with YV2 RVs. Once again, we ran a MCMC as in Sect.~\ref{sec:hd192310} to get a new updated Keplerian solution for the system and updated minimum masses. The orbital parameters are displayed in Table.\ref{Table3}. Using the YV2 Keplerian solution, the residual RV rms was decreased from 1.46\,\ms{} with the DRS dataset down to 0.93\,\ms{} after YV2.

The system published by \citet{Vogt(2010)} is now recovered and the RV rms is below 1\,\ms{}. This example shows once again how a 1-year systematics can mix with planetary signals to produce a peak elsewhere in a periodogram (often at the 1-year alias of the real signal), an element already raised in \citet{Cretignier(2021)}. This phenomenon is more likely to happen when the systematics and the underlying signals are comparable in amplitude which is the case here. 

\begin{figure}
	\centering
	\includegraphics[width=9cm]{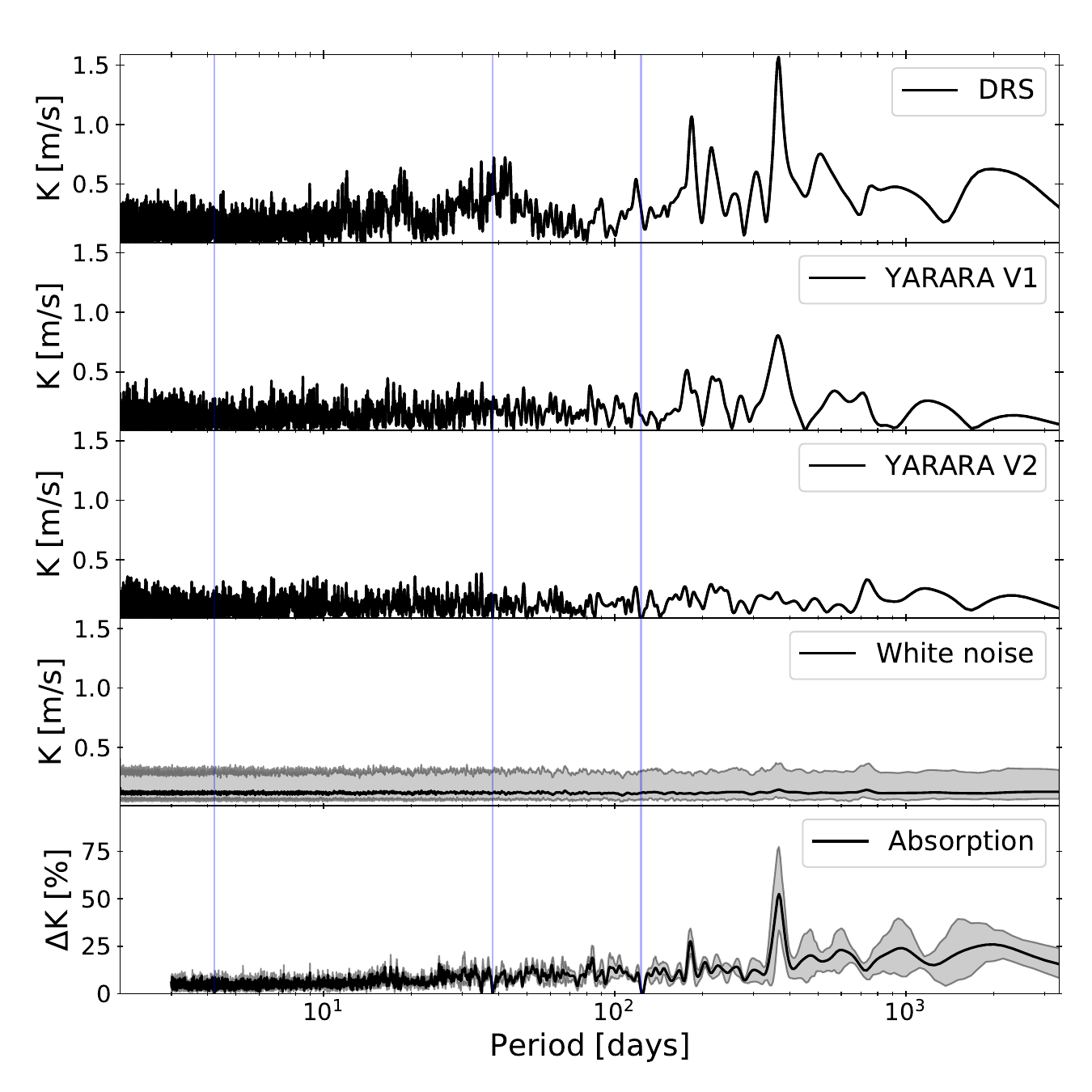}
	\caption{Same as Fig.~\ref{FigGlobalHD192310} for HD115617. The Keplerian solutions obtained with the YARARA V2 dataset was removed. The absence of power at 4, 38 and 120 days in all the datasets shows that the Keplerian solution used is a valid solution for all of them. Clear systematics at 365 days and 182 days are visible in the DRS time-series (first row), mainly created by the stitching effect \citep{Dumusque(2015b)}. In YV1 RV time-series (second row). Power excess around 33 days was strongly mitigated as well as the previous 1-year and first harmonic systematics. Remaining 1-year power is however still visible and is only fully corrected in the YV2 RV time-series (third row) that exhibits a periodogram compatible with a white noise signal (fourth row). This 1-year correction is explained by the strong ability of the basis to absorb 1-year signal (up to 75\% absorption), as visible in the absorption curve (fifth row).}
	\label{FigGlobal}
	
\end{figure}

To further confirm it, we display in Fig.~\ref{FigGlobal} the periodogram of the residuals RV datasets with the Keplerian solution obtained by the MCMC on YARARA V2. Even if the 1-year power is not visible when fitting iterative Keplerian signals (as showed in Fig.~\ref{Fig61Vir1}), the signal is clearly visible on the DRS dataset with a semi-amplitude of $K=1.50$ \ms, likely dominated by the stitching \citep{Dumusque(2015)}. Its strength is strongly reduced after YARARA V1 down to $K=0.70$ \ms, but fully disappear only after YARARA V2 processing. The absence of peak at the planetary periods shows that the Keplerian solution used is also working for all the datasets which confirms that signals were not absorbed in YV2.

Even if a similar system is now recovered with HARPS than in \citet{Vogt(2010)}, we noted a major difference about the minimum mass of 61\,Vir, d. Indeed, the planetary mass published for that planet is about $m \sin i=24.0$ \Mearth{} which is a massive Neptune-like planet whereas in our case, the planetary mass detected is around $m \sin i=10.5\pm0.8$ \Mearth{}, hence close to half the mass of Neptune which is rather the massive range of Super-Earth regime. The planetary simulations performed on HD10700 (Sect.~\ref{sec:hd10700}) prevents the possibility from an absorption of the signal coming from the YARARA reduction which is also confirmed by the similar amplitude obtained for the DRS dataset. As a matter of fact, the amplitude is also in agreement with \citet{Laliotis(2023)}. 

It is known that instrumental systematic lead to biased eccentric Keplerian solutions \citep{Hara(2019)} with mass overestimated, the eccentric orbits allowing to absorb part of the systematics. This effect is visible on our case by the large and unconstrained eccentricity $e=0.17^{+0.31}_{-0.12}$ of the planet $d$ on the DRS dataset that reduces to $e=0.12^{+0.05}_{-0.05}$.

We did not try to include the published HIRES RVs with the present corrected HARPS RVs since we do not know the red noise model related to HIRES. Instead, we investigated the residuals of the HARPS Keplerian solution on the HIRES RV time-series visible on DACE, similarly to the analysis with the DRS dataset in Fig.~\ref{FigGlobal}. No significant peak at the planetary periods remains in the periodogram of the HIRES residual time-series, demonstrating that the present Keplerian solution is also working with their instrument. 

The present example is crucial to understand the difficulty to obtain accurate mass measurement in the presence of red noise improperly modeled or for signal close to the instrumental precision limit. As a matter of fact, future space PLATO mission would require mass measurement accurate at 20\% \citep{Rauer(2016)}, and similarly for the interpretation of JWST observations \citep{Batalha(2019)}, which can already be challenging in the Super-Earth regime ($m_p$<10 \Mearth{}) and even more for the Earth-like planetary regime ($m_p$<5 \Mearth{}).

\begin{figure*}
	
	\centering
	\includegraphics[width=17.5cm]{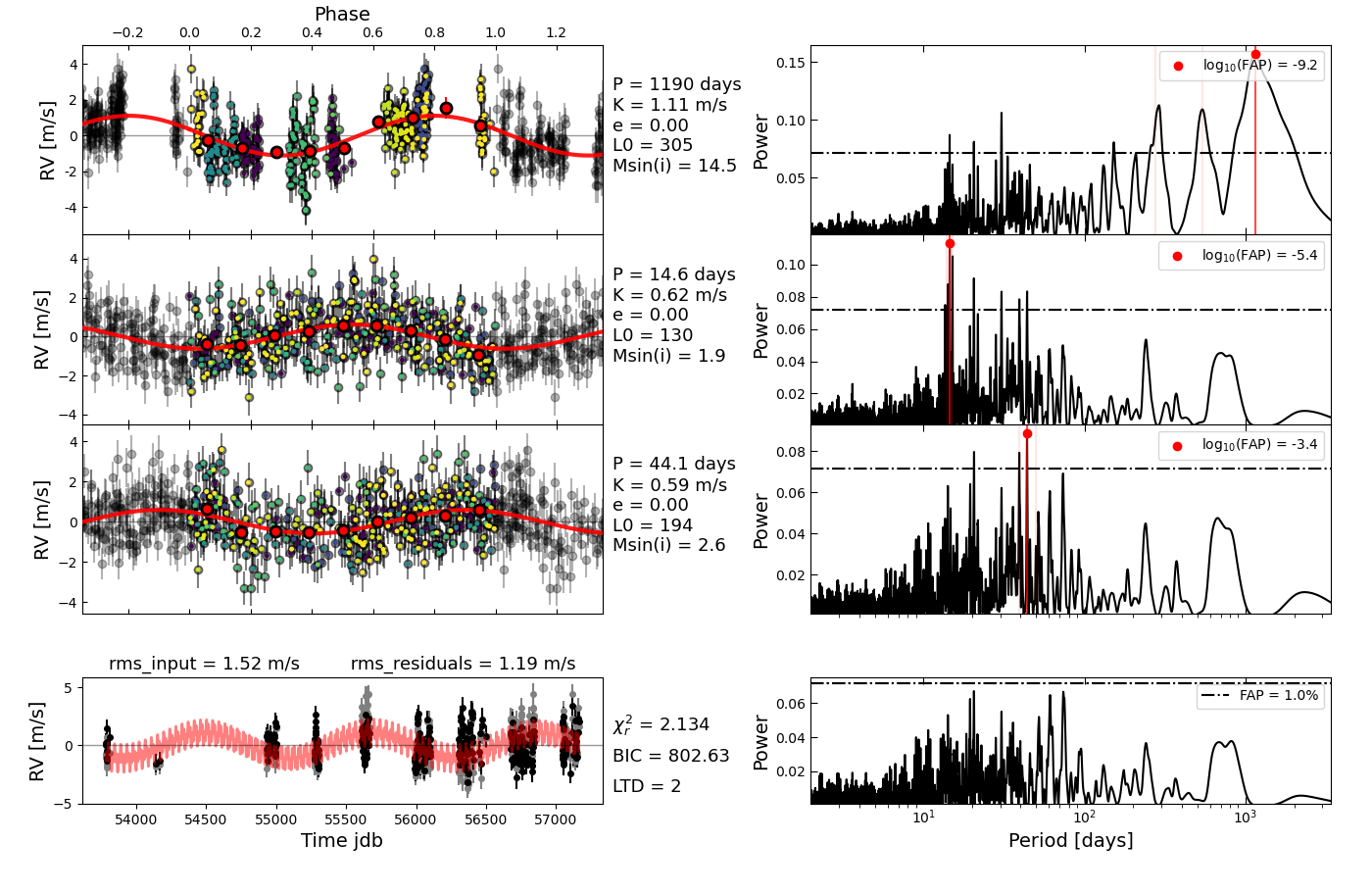}
	\caption{
	Iterative circular orbits Keplerian-fit of HD109200 with the RV time-series of the DRS.}
	\label{FigHD1092001}
	
	\centering
	\includegraphics[width=17.5cm]{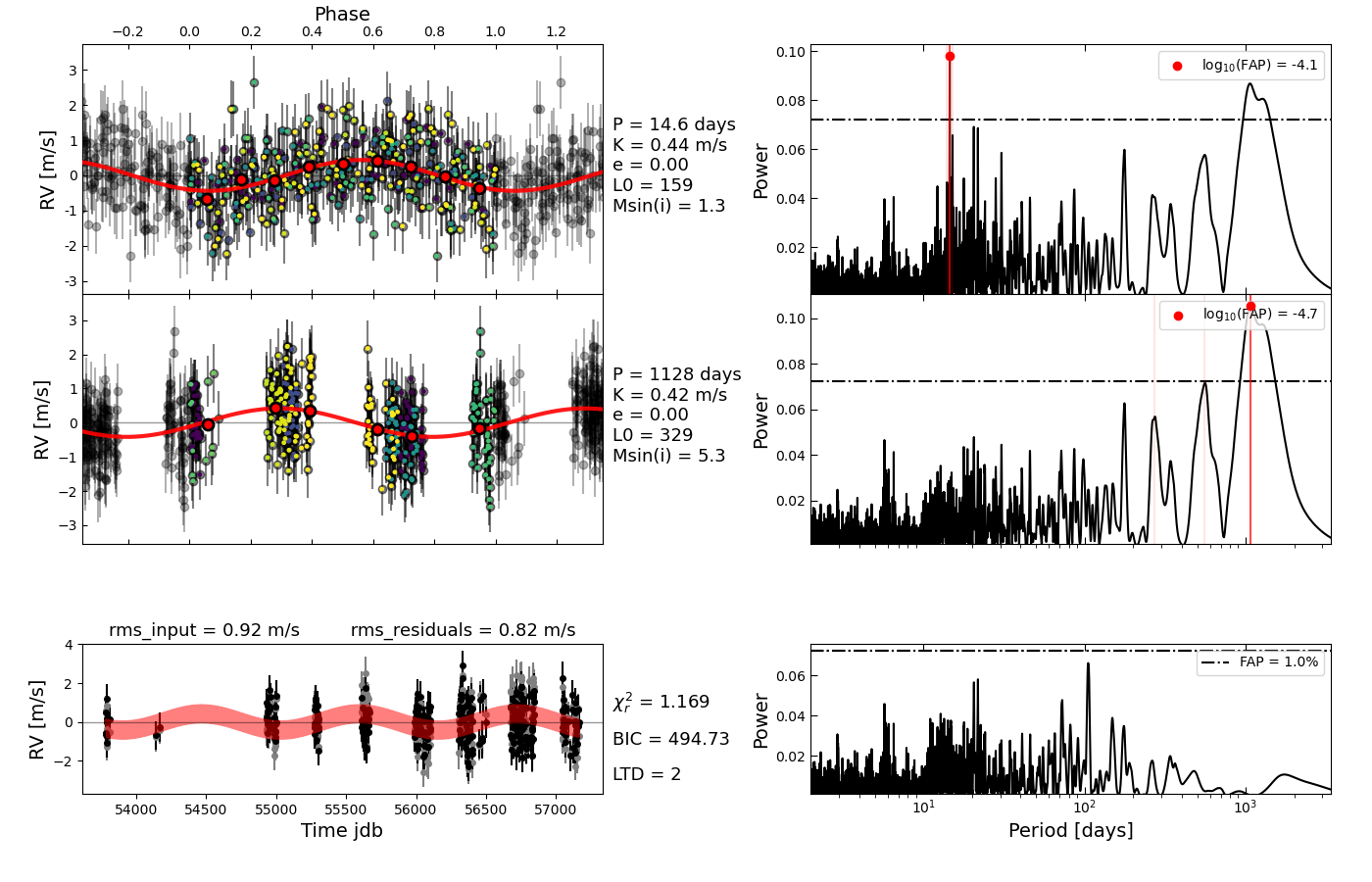}
	\caption{
	Same as Fig.~\ref{FigHD1092001} with the RV time-series from YARARA V2.}
	\label{FigHD1092002}
	
\end{figure*}

\subsection{HD109200}
\label{sec:hd109200}

HD109200 is a rather bright ($m_v=7.1$) K1V star located at 16.2 pc of the Sun with a stellar mass of 0.70 \Msun{} \citep{Sousa(2008)}. The present dataset has the lowest S/N of the paper with $med(S/N)=189$. In total, 357 observations were taken between 17th February 2006 and 21th May 2015. The star shows a very irregular magnetic cycle with a sudden outburst of activity during the 2011 season. There are a few evidences of a rotational period between 36 and 40 days based on chromospheric activity proxies.

When fitting iterative circular orbits (see Fig.~\ref{FigHD1092001}), at least three signals are detected at 15, 42 and 1200 days. For that dataset, the improvement is less noticeable than with the other ones. After YV1, the 42-day signal disappears, whereas the 1200-day signal decreases down to 55 \cms. That amplitude is further reduced down to 40 \cms after YV2 (see Fig.~\ref{FigHD1092002}) whereas the 15-day signal remained unchanged. In YV2, only 7 components were detected as significant (5 \textit{shell} and 2 \textit{slice}). This effect is likely due to the lower S/N of the observations. Indeed, in \citet{Cretignier(2022)}, the authors already showed that, below S/N$<250$, data-driven approach, such as the "shell" framework, may drastically decrease in performance. The present example further confirms that aspect since no component was detected as significant, neither on CBC RVs nor LBL RVs, producing a smaller vector basis to decorrelate the RVs in YV2.

\begin{figure}
	\centering
	\includegraphics[width=9cm]{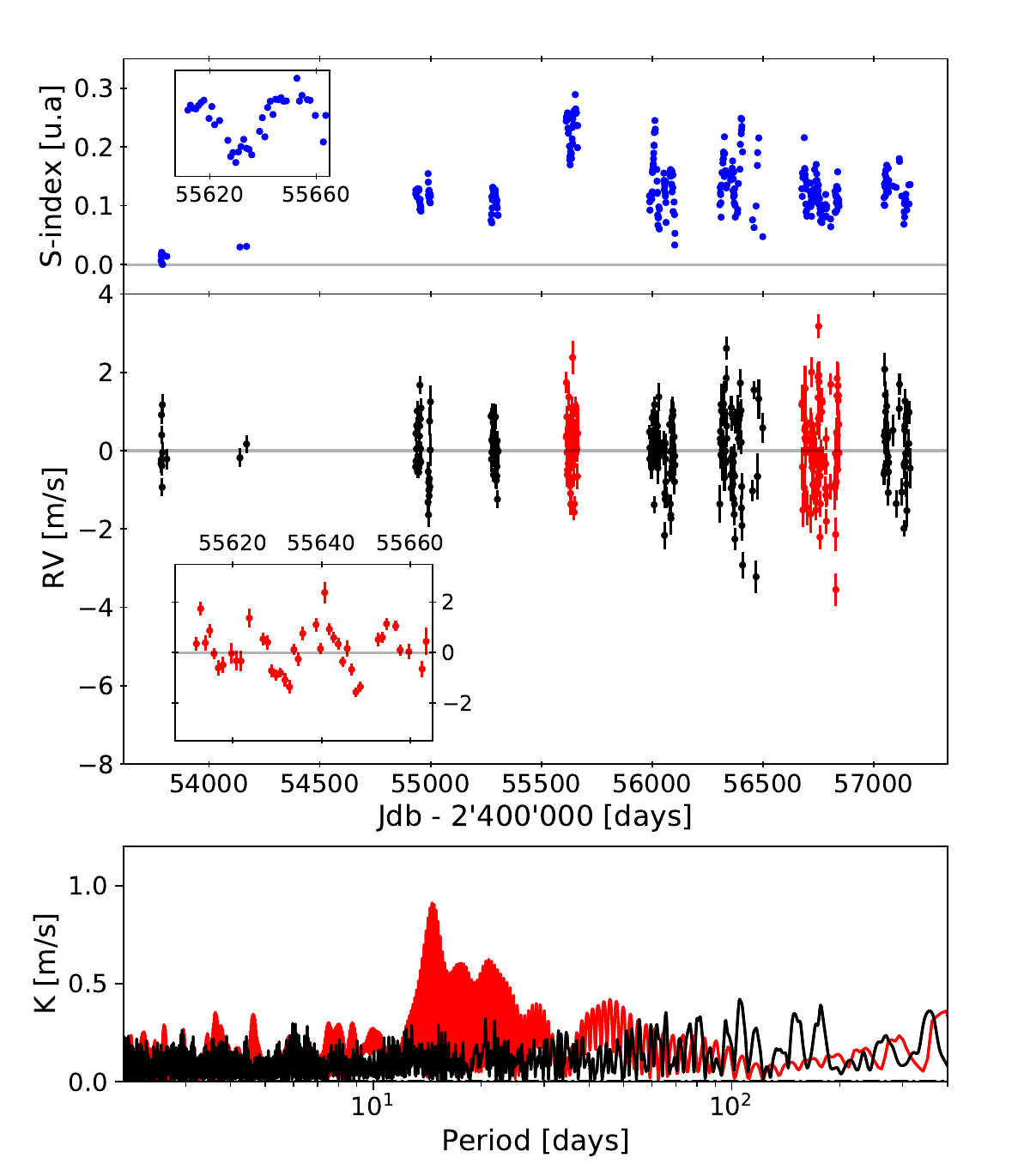}
	\caption{Analysis of the spurious 15-day RV signal on HD109200. \textbf{Top panel:} S-index time-series. The star exhibits an irregular magnetic cycle with a sudden outburst of activity in 2011. The inner panel is a zoom on the 2011 season where the 36-day rotational period is visible. \textbf{Middle panel:} YV2 RV time-series. Each observational season has been mean-centered to remove the long trend signal. The inner panel shows a zoom on the 2011 season for which the 15-day signal is clear. \textbf{Bottom panel:} GLS periodogram of the 2011 and 2014 season (red curve) and of the remaining time-series (black curve). The 15-day signal with an amplitude of $K=1.0$ \ms is exclusively coming from those two seasons.}
	\label{FigHD109200}
	
\end{figure}

None of the remaining signals are convincing and could pretend to become planetary candidates. All failed the \textit{quadrature attack} test and their periods were therefore removed from the model since they were initially added by the l1-periodogram. We investigated the 15-day signal and remarked that the power is almost exclusively coming from the 2011 and 2014 seasons. This is shown in Fig.~\ref{FigHD109200} where we displayed the YV2 time-series periodogram on those two seasons and on the full time-series excluding those seasons. For this analysis, we also mean-centered each observational season to remove the long trend signal. Whereas a $K=1.0$ \ms signal is observed around 15 days on the dataset made of 2011 and 2014, nothing is visible on the remaining seasons. 

The origin of the 15-day signal is unclear. A stellar activity origin may be expected given that 2011 is the season with the highest activity-level but 15 days is not a classical harmonic of any period between 36 and 40 days which is the rotational period range expected for the star. However, we noted that a similar periodicity of 15 days is found in the VSPAN of the CCF, but a lag of $30^{\circ}$ exists between both signals. This may indicate that the RV signal is mainly induced by spots rather than facula in 2011 since spot tend to present a larger VSPAN signature than facula due to their larger contrast, whereas the RV signal of faculae correlated better with the S-index which is not the case here. This may also explain why the shell framework did not correct for it since the methodology was developed to correct the inhibition of the convective blueshift from facula and not the flux effect from spots.

For the 1200-day signal, its amplitude is strongly reduced from the DRS to YV2. Furthermore, the lack of significant components from PCA on CBC RVs or LBL RVs is concerning since 1-year systematics are usually corrected by those recipes. As a last concern, the signal is poorly sampled in phase and the baseline only cover two orbital phases, which is insufficient to validate a 50 \cms signal. This argumentation is particularly motivated given that a 1400-day signal is rather fit if no long-trend drift is fit in the model. For this reason, we also consider this signal as unreliable at the moment. 

Note that despite a clear irregular magnetic cycle and even if two doubtful signals remains, the intrinsic precision of the star around 80\,\cms{} is one of the best achieved on the HARPS program.

With this example, we highlight the difficulty to obtain significant components from PCA on dataset with S/N < 200, even if many observations exists ($N>350$). Not only the quantity of the data, but also their quality matters. Because of it, some signals that are likely not planetary by nature are founded robust to the approach presented in this paper. Some improvements can still be observed since part of the signals disappear and since the RV rms decrease from 1.19 to 0.92 \ms, but the improvement is far less notable compared to the other dataset.

We also highlight that HD109200 appears as a really interesting dataset for Extreme Precision RV (EPRV) challenge as the ones performed recently \citep{Dumusque(2016),Zhao(2022)} that aims to compare the current methods of stellar activity mitigation and RV extraction.

\subsection{HD20794}
\label{sec:hd20794}

HD20794 is a bright ($m_v=4.3$) G6V star located at 6.1 pc of the Sun with a stellar mass of 0.81 \Msun{} \citep{Ramirez(2013)}. The star didn't show any magnetic cycle along its lifetime and there is no hint of a rotational modulation period based on chromospheric activity proxies.

\begin{figure*}
	
	\centering
	\includegraphics[width=17.4cm]{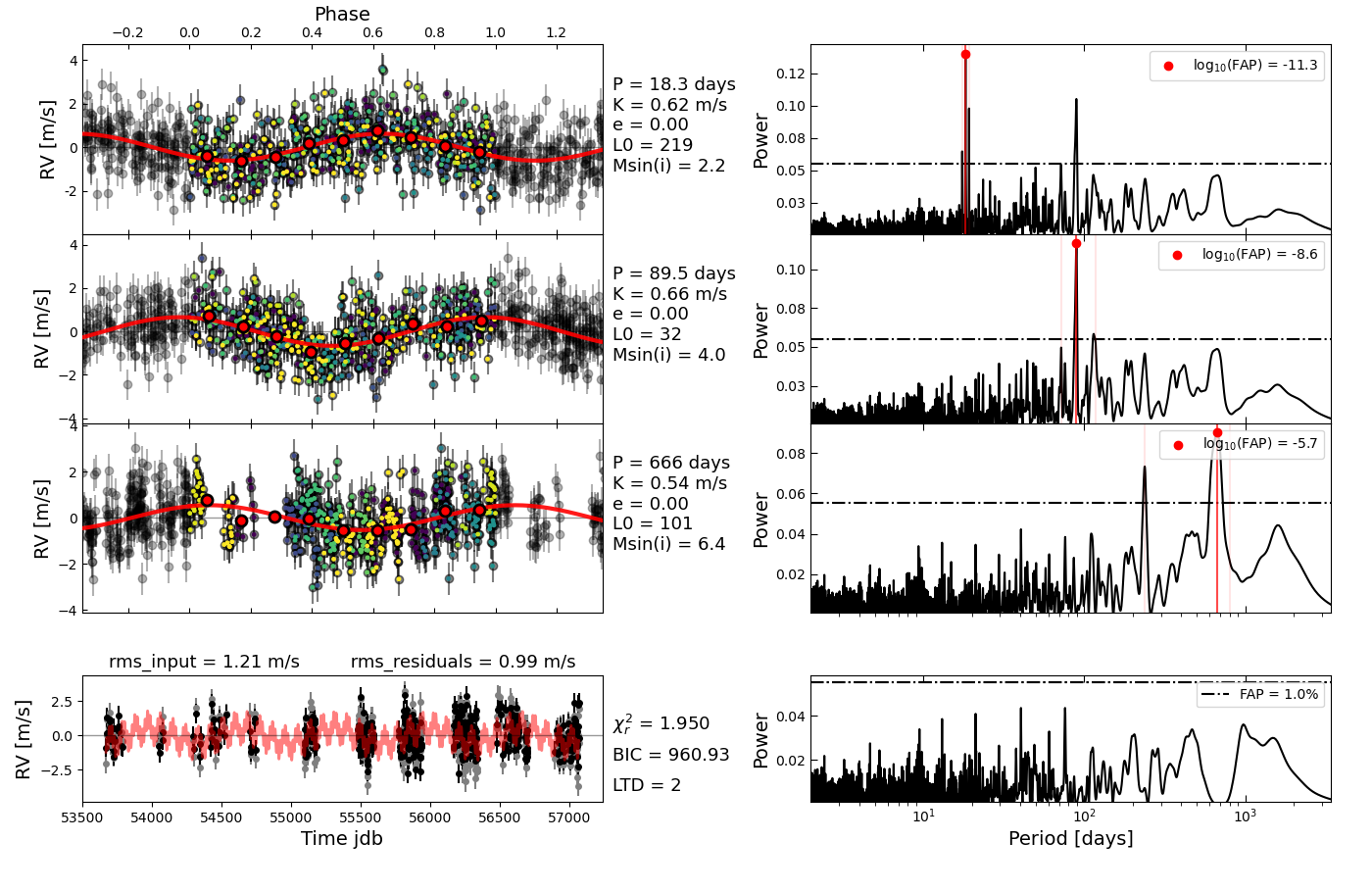}
	\caption{
	Iterative circular orbits Keplerian-fit of HD20794 with the RV time-series of the DRS. The two shortest signals are already published.}
	\label{FigKep1}
	
	\centering
	\includegraphics[width=17.4cm]{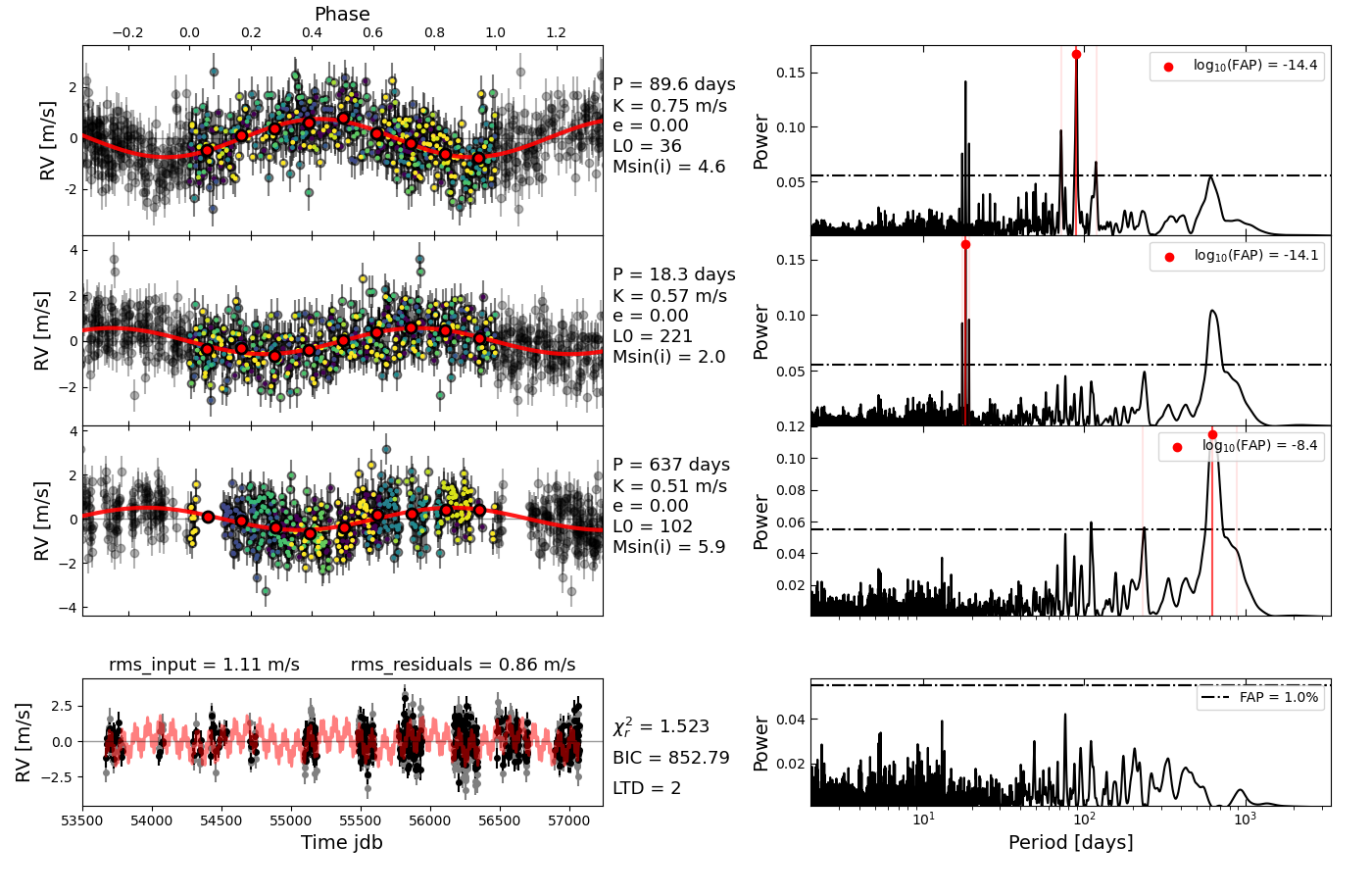}
	\caption{
	Same as Fig.~\ref{FigKep1} with the RV time-series from YARARA V2. The significance of the 650 days signal is strongly boosted.}
	\label{FigKep2}
	
\end{figure*}

The star has been intensively observed during the full lifetime of the instrument with 468 nightly measurements between $25^{th}$ October 2005 and $13^{th}$ February 2015 and a rms of 1.36\,\ms{} of the same baseline. The median S/N of the dataset is the largest one for this work with a value of $med(\text{S/N})=459$. Several teams already proposed different candidates for that systems \citep{Pepe(2011b),Feng(2017)} where only partial agreements were found for a 18 and 89 days signals. When analysing the RV coming from the DRS, two clear signatures are detected at 18 and 89 days (also found in the previous mentioned studies) attributed to two Super-Earth exoplanets in a case of an equator-on system with $K$ semi-amplitudes of 60\,\cms{}. The DRS dataset also provides a longer signal at 650 days that could be attributed to another super-Earth planetary candidate (see Fig.~\ref{FigKep1}). 

After YV1, the 650-day signal is replaced in the iterative GLS by a 410-day and 240-day signals. However, when fitting a l1-periodogram, the system detected is rather a 650-day signal with a 1-year signal. We notice that 650 and 240 days are 1-year aliases of each other, and we therefore interpret this result as a cross talk between the yearly signal and the 650 day signal, as in the GLS approach, contrary to the l1-priodogram, we fit signals iteratively. We further confirm that the 650-day signal appears as the good planetary candidate since the l1-periodogram using the shell and PCA components also converge to a solution of three signals at 18, 89 and 650 days. Since the 650-day was detected in the l1-periodogram, this latter was added into the Keplerian solution as described in Sect. \ref{sec:keplerians}.

On YV2, 14 components were fit (5 \textit{shell}, 2 \textit{slice}, 3 \textit{color}, 4 \textit{lbl}), and the system recovered now contained the 650-day which is boosted to $\log_{10}(FAP) = -8.4$ (see Fig.~\ref{FigKep2}). To check the reliability of the signal, we tried to remove it from the $\text{RV}_P(t)$ basis to see if its power vanished. Even if, this latter was indeed reduced due to the long-trend absorption ability ($\Delta K\sim30$\%), the signal was clearly visible and significant with a $\log_{10}(FAP) = -2.9$. This further confirm the robustness of the signal as a strong potential planetary candidate. With the \textit{quadrature attack} test, we noticed that the power of the signal comes closer in amplitude to its 1-year alias at 240-day, but still remains above it.  

The Keplerian parameters obtained with the MCMC are written in Table.\ref{Table3}. The residuals RVs from the MCMC solution is 0.85\,\ms{} for YARARA V2. The $K$ semi-amplitude is found to be a $10\sigma$ detection of a 60\,\cms{} amplitude signal corresponding to a planetary minimum mass of $m\,\sini{}=6.6\pm0.7 \Mearth{}$ orbiting at $1.36\pm0.03$ AU. We notice that the eccentricity of the planet $c$ was decreased whereas the eccentricity of the 650-day planetary candidate is rather fixed around $e=0.40\pm0.07$. 

\begin{figure}[t]
	
	\centering
	\includegraphics[width=9cm]{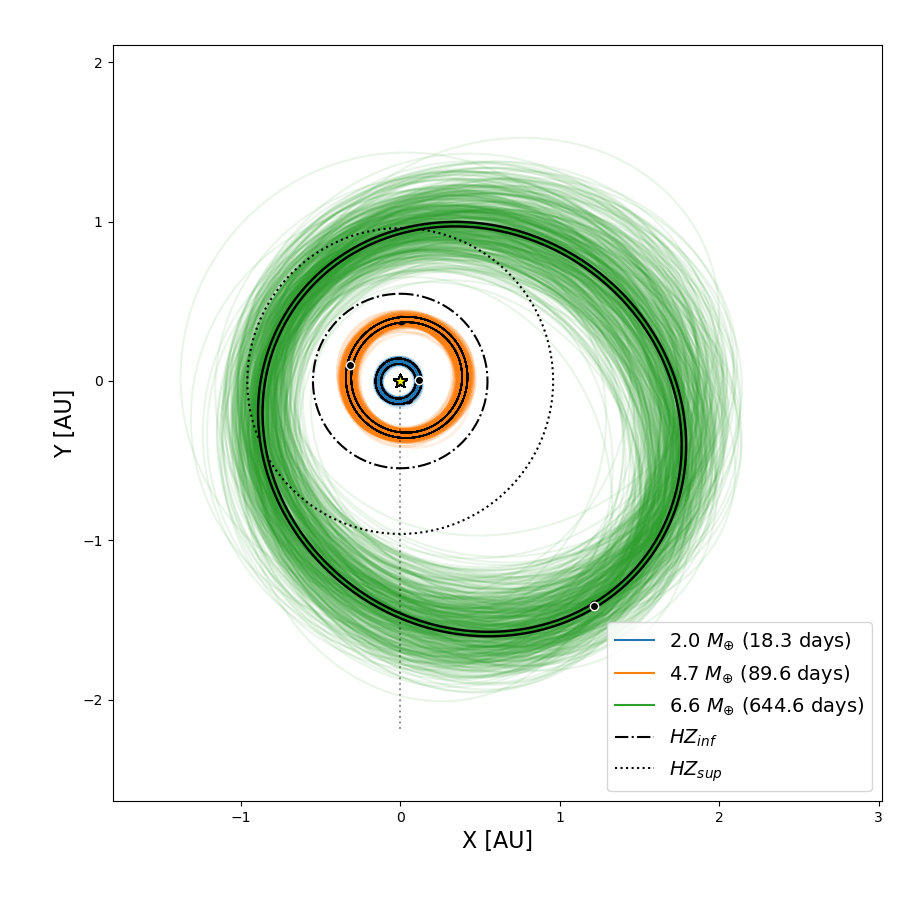}
	\caption{Resulting MCMC simulations trajectory for a 3-Keplerian fit model (co-orbital configuration) obtained with the YARARA V2 RV time-series of Table.\ref{Table2}. The stellar mass of HD20794 is taken as $M_*=0.81$ \Msun{}. If confirmed, the third signal at 644 days would orbit, during part of its orbit, in the habitable zone of the star \citep{Kopparapu(2013)} between the innermost (dotted-dashed circle) and outermost limit (dotted circle). The orbital solutions from the median values of the posterior distributions of the Keplerian parameters are also displayed (main solid lines).}
	\label{FigMCMC}
	
\end{figure} 

Interestingly enough, such a high eccentricity would have a peculiar property on the system if confirmed. As displayed in Fig. ~\ref{FigMCMC}, the planetary orbit crosses the habitable zone (HZ) of the G6V star at its apsis, which would imply that seasons on this potential exoplanet could be induced by the eccentricity of the orbit rather than the planetary obliquity. To visualize it, we represent the orbital solutions from the MCMC as well as the orbital solutions coming from the median of the posterior distribution and the inner and outer habitable region according to \citet{Kopparapu(2013)}. Nevertheless, even though the MCMC converges to a clear eccentric solution, small unknown systematics may still be responsible for the large eccentricity and a circular orbit solution (as the one obtained in Fig.~\ref{FigKep2}) still looks as a very good fit which would be favoured based on BIC criterion. In such a case, the candidate would orbit outside the HZ. 

This signal couldn't be detected yet in \citet{Pepe(2011b)} due to the shorter baseline of the observations and the unknown existence of the long-term instrumental systematics. Whereas this signal was mainly undetectable in \citet{Feng(2017)} due to the wrong assumptions about the instrumental noise model used by the authors, assumed to be mainly short timescale jitter. Indeed, we demonstrated in \citet{Cretignier(2021)}, as well as in the present paper, that instrumental systematics also contaminate at longer periods ($P>300$ days). This third signal is strongly convincing, but we are aware how difficult it could be to confirm it given the semi-amplitude and periods of the signal. It raises the question of how such candidates could be validated? It is likely that several new extra years of observations and hundreds of new nights would be necessary to confirm such a small signal with RVs. 

\begin{figure}
	\centering
	\includegraphics[width=9cm]{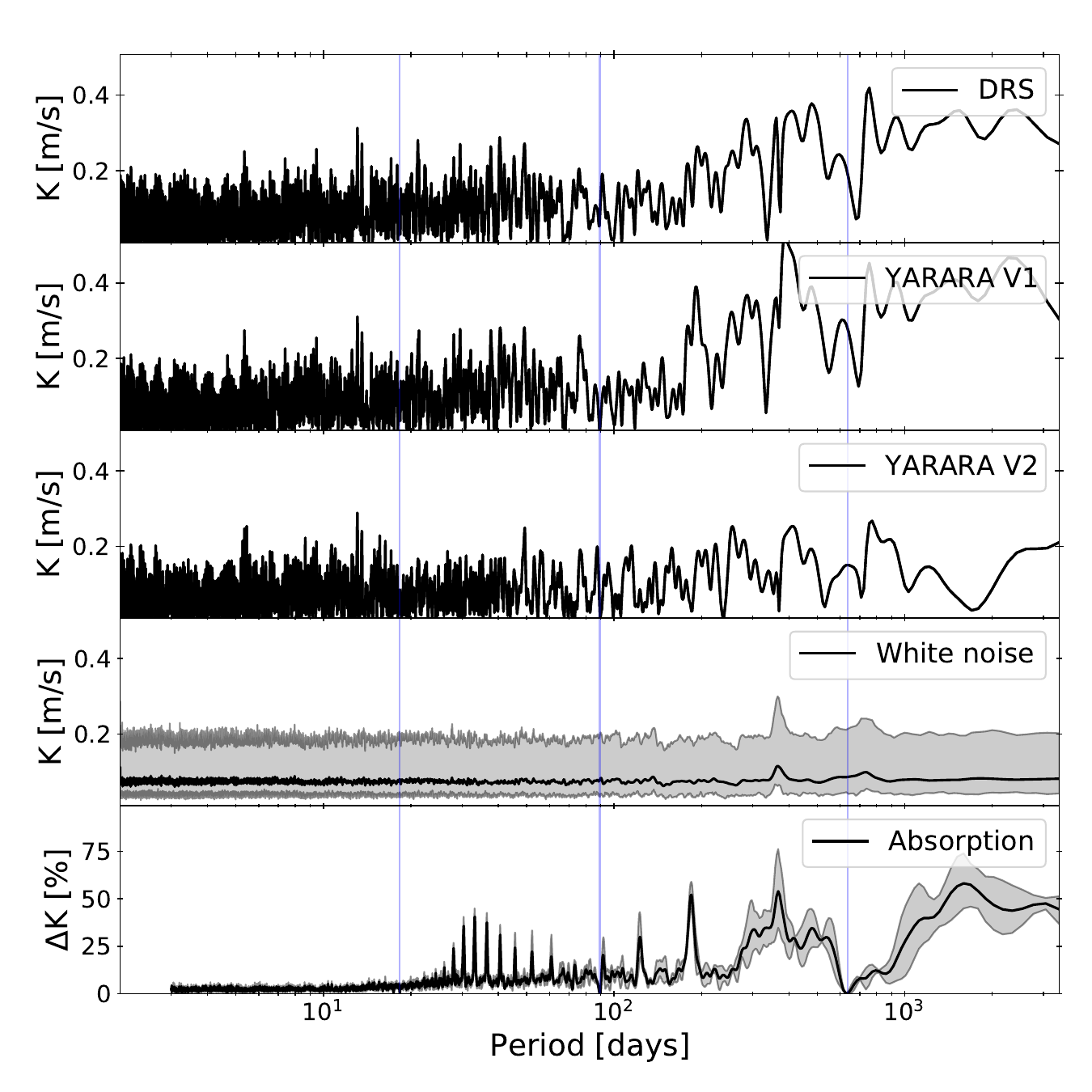}
	\caption{Same as Fig.~\ref{FigGlobalHD192310} for HD20794. The Keplerian solutions obtained with the YARARA V2 dataset was removed. A comb of absorption signal is visible 35 days and is due to the interference pattern (see Appendix \ref{appendix:b}). The long-trend absorption is also clearly visible.}
	\label{FigGlobal3}
	
\end{figure}

In Fig.~\ref{FigGlobal3}, we show that no  signal amplitude larger than 25\,\cms{} is visible in the residuals YV2 RV time-series. We also highly that the absorption curve exhibit a comb structure due to the new detected 0.1 \ang{} interference pattern that is only detectable in LBL PCA of bright targets. Hopefully, the forest of peaks for this star is centered around 32 days and therefore falls between the two planetary candidates at 18 and 90 days. The PCs produced by the interference pattern (see Appendix \ref{appendix:b}) are the most risky components encountered until now since they can lead to absorption up to $\Delta K =50\%$ for periods matching the interference frequency. As a consequence, interference patterns may be a strong limitation for PCA on LBL RVs.

We note that further analyses should be conducted to confirm or infirm the present new planetary candidate here proposed, since part of the phase is yet uncovered due to seasonal gap. Furthermore, without any knowledge about the inclination angle of the system, the Super-Earth nature of the signal is not yet even determined. Recalling that a factor larger than two in mass ($i>60^\circ$) happens for 13\% $\sim \frac{1}{6}$ of the exoplanets for an isotropic \sini{} distribution which is not so unlikely. This is particularly true when thinking that the rotational period of the star is not known and detecting the rotational period of pole-on stars is more challenging than equator-on ones \citep{Borgniet(2015)} due to the smaller \vsini{} value and that active region remain located to the stellar limb\footnote{Assuming a Sun-like butterfly diagram.}.

\renewcommand{\arraystretch}{2.0}
\begin{table*}[tp]
\caption{Orbital and physical parameters of the planets obtained from the MCMC performed on the YARARA V2 RV time-series of HD20794 for a 3-Keplerian fit model. The reference date is $BJD=2'455'500$ and the stellar mass is taken as $M_*=0.81$ \Msun{} \citep{Ramirez(2013)}.}
\label{Table2}
\centering
\begin{tabular}{c|ccc|ccc|ccc}
\hline\hline
 & & DRS & & & YV1 & & & YV2 & \\
\hline
Par. & planet b & planet c & planet d & planet b & planet c & planet d & planet b & planet c & planet d \\
\hline
$P$ & $18.32^{+0.01}_{-0.01} $ & $89.39^{+0.09}_{-0.09} $ & $654.4^{+7.9}_{-4.5} $ & $18.32^{+0.01}_{-0.01} $ & $89.55^{+0.10}_{-0.10} $ & $701.9^{+5.6}_{-6.4} $ & $18.32^{+0.01}_{-0.01} $ & $89.58^{+0.09}_{-0.10} $ & $644.6^{+9.9}_{-7.7} $ \\
$K$ & $0.60^{+0.05}_{-0.05} $ & $0.82^{+0.08}_{-0.07} $ & $0.65^{+0.07}_{-0.08} $ & $0.59^{+0.05}_{-0.05} $ & $0.76^{+0.08}_{-0.07} $ & $0.66^{+0.07}_{-0.07} $ & $0.56^{+0.05}_{-0.05} $ & $0.78^{+0.05}_{-0.05} $ & $0.61^{+0.06}_{-0.06} $ \\
$e$ & $0.06^{+0.07}_{-0.04} $ & $0.45^{+0.07}_{-0.09} $ & $0.37^{+0.07}_{-0.08} $ & $0.10^{+0.09}_{-0.07} $ & $0.36^{+0.10}_{-0.10} $ & $0.55^{+0.07}_{-0.07} $ & $0.09^{+0.08}_{-0.06} $ & $0.13^{+0.07}_{-0.07} $ & $0.40^{+0.07}_{-0.07} $ \\
$\omega$ & $325^{+112}_{-83} $ & $170^{+8}_{-9} $ & $201^{+15}_{-17} $ & $293^{+69}_{-46} $ & $162^{+11}_{-18} $ & $115^{+15}_{-12} $ & $349^{+60}_{-61} $ & $151^{+29}_{-36} $ & $214^{+16}_{-19} $ \\
$\lambda_0$ & $220^{+5}_{-5} $ & $35^{+4}_{-4} $ & $94^{+6}_{-5} $ & $219^{+5}_{-5} $ & $38^{+4}_{-4} $ & $99^{+7}_{-7} $ & $222^{+6}_{-5} $ & $36^{+4}_{-4} $ & $102^{+7}_{-5} $ \\
\hline
$a$ & $0.13^{+0.01}_{-0.01} $ & $0.36^{+0.01}_{-0.01} $ & $1.38^{+0.03}_{-0.03} $ & $0.13^{+0.01}_{-0.01} $ & $0.37^{+0.01}_{-0.01} $ & $1.44^{+0.02}_{-0.03} $ & $0.13^{+0.01}_{-0.01} $ & $0.37^{+0.01}_{-0.01} $ & $1.36^{+0.03}_{-0.03} $ \\
$m$ \sini{} & $2.1^{+0.2}_{-0.2} $ & $4.4^{+0.4}_{-0.4} $ & $7.1^{+0.8}_{-0.8} $ & $2.1^{+0.2}_{-0.2} $ & $4.3^{+0.4}_{-0.4} $ & $6.6^{+0.8}_{-0.8} $ & $2.0^{+0.2}_{-0.2} $ & $4.7^{+0.4}_{-0.4} $ & $6.6^{+0.6}_{-0.7} $ \\
\hline
\end{tabular}
\end{table*}

The present example was more dedicated to illustrate three conclusions: 1) other planetary candidates signals (18 and 90 days) are plainly recovered with similar amplitudes after YARARA V2 than with the DRS dataset and their significance are increased, 2) we reconfirm the stability of the HARPS instrument below 1\,\ms{} on a baseline of 10 years with residuals RVs rms of 86\,\cms{} (similarly as HD10700 in \citet{Cretignier(2021)} and 3) looking at the high FAP value of the planetary candidate, the detection of Super-Earth exoplanets in the habitable zone of solar type-stars is already possible on intensive program of observations. As a consequence, the application in the future of YARARA V2 on historical archived database could lead to unveil or confirm promising planetary candidates.

\section{Conclusion\label{sec:conclusion}}

We presented in this paper a refined version of our previous YARARA methodology where further corrections can be performed on LBL RVs thanks to PCA. The PCA appears as a well dedicated tool to detect systematics similarly to its recent use in \citet{Ould(2023)}. Its mean-invariant property allows the PCs to be insensitive to planetary signals. In this paper, we managed to detect and fit for the ThAr ageing using only stellar observations. To our knowledge, this is the first time that such corrections is performed without calibrations products. 

A first exploratory approach allowed us to notice that several HARPS targets were sharing similar PCs. We therefore developed some Z-score calibrations curves in order to average the LBL RVs and reinforce the signal along the systematic variance direction. From these residuals LBL RVs, lines were then averaged in chunk-by-chunk of 4 \ang{} in order to highlight smooth chromatic effects. Finally, a last PCA was finally ran on the individual LBL RVs for a free correction. 

We tested our methodology on five stars intensively observed by HARPS. We demonstrated on them that the planetary signals (either injected or real ones) were perfectly recovered and we reached for all the targets RV residuals rms around 90\,\cms{} on HARPS, for a baseline up to 10 years for the most observed ones. Given that some PCA vectors are clearly related to instrumental systematics and wavelength solutions issue (such as the ThAr component), we could wonder if $\sim$80\,\cms{} is not an intrinsic limit difficult to overcome because of granulation or supergranulation signals as raised in \citet{Meunier(2015),Meunier(2020),Almoulla(2023)}. This is particularly true given that our best targets reach hardly better than 80\,\cms{} whereas the periodogram of the residual RVs does not exhibit any convincing peak that may be hidden planetary signals or remaining red noise. Such a conclusion may have a huge consequence for the number of nights or the baseline of long monitoring RV surveys  \citep{Thompson(2016),Hall(2018),Gupta(2021)}. However, we found compatible RV residual rms for stars with activity cycle (HD192310 and HD109200) compared to stars without clear cycle (HD20794). As a final remark, detecting lower semi-amplitude signal than $K<80$ \cms is possible and likely achievable with RVs (as showed in this paper with HD20794), but simply require a large number of nights.  

On HD192310, we demonstrated that the rotational period of the star was no more significant after YARARA V2 leading to a similar conclusion that in our previous paper for $\alpha$ Cen B \citep{Cretignier(2022)}. The mass of the most external planet was however found 40\% lower than the published one. 

Around HD115617, a system of three exoplanets already published by HIRES RVs measurements, the expected system is recovered opposite to the DRS case that was converging to a wrong system due to crosstalk between the signals and the systematics. Similarly, we found the mass of the 120-day planet strongly overestimated by the HIRES team, which can be explained by the low semi-amplitude of the signal ($\sim$1.5\,\ms{}) already below the intrinsic stability of the instrument. Since the signal was debated in the literature, we confirm that this one is likely a real planetary signal in agreement with \citet{Laliotis(2023)} and in disagreement with \citet{Rosenthal(2021)}.

For HD109200, we showed some limitations with LBL PCA to deal with moderated S/N dataset (S/N$<200$). Even if a slight improvement can be observed and a signal around the rotational period disappear, a clear non-Doppler signal at 15-day is still detected. In particular, we point out that the 2011 observational season may be considered as an excellent dataset to test the correction ability of different methods. 

Finally, around HD20794, the two inner planetary candidates were boosted and a 650-day signal slightly significant was strongly boosted after systematic corrections. The signal could correspond to a Super-Earth crossing the habitable zone of the star during its revolution, even though a circular orbit outside the outerbond of the HZ could fit the data as good as the eccentric solution. With the current RV precision, it is difficult to conclude on the nature of the signal and a few hundreds more observations over at least 2 years could help in evaluating the planetary nature of this candidate. We further notice that such candidate would be one of the closest Earth-twins discovered so far.

It should be reminded that the improvement of the RV precision here obtained by YARARA V2, on the long-term by correcting instrumental systematics and short-term with stellar activity, were all obtained by a new methodology of planetary-free proxies' extraction based on new representations of the EPRV problem. At the end, the model fit by YARARA V2 is no more complex than a multi-linear regression of a dozen of vectors, which is a small number of degrees of freedom compared to the expected number of visits for the dataset for which YARARA has been designed to work. Because of their rigidity and simplicity, multi-linear models are less computationally expensive than kernel regression or Gaussian Process and are less likely to absorb real planetary signals. GPs often require a good enough understanding of the red noise component (through the definition of a covariance matrix) which is difficult to obtain for instrumental systematics. Moreover, since PCA are by nature multi-linear decomposition, using such a model to correct the RVs themselves makes sense.

Further improvements can be performed, namely by the production of other $Z-$score calibration curves for the other remaining systematic or for the stellar activity which would help such methodology to be deployed on less bright stars in a more efficient way. Also an ultimate solution to avoid any cross-term effect would be the intrinsic knowledge about the $\beta_{i,j}$ coefficients, avoiding in that way the problematic simultaneous fit with the Keplerian signals. But it would require a far better understanding of the systematics here corrected which is not an easy task to handle. At this regard, the new proxies delivered by YARARA V2 could be investigated in order to better understand their origin or their physical interpretation. This was for instance used to flag lines that were the most affected by the new detected interference pattern, which allowed to visually detect it afterwards (Appendix \ref{appendix:b}). As a final warning, we point out that even though LBL PCA appears as a powerful new generation tool, the difficulty and limitations that interference pattern introduce in the PCA components should be understood and such pattern should be avoided as much as possible at the instrumental level. 

Finally, we also highlight that such a strategy of PCA decomposition can be applied on any series of RV time-series. Another application could be the order-by-order CCF RV time-series, which could be useful for dataset with lower S/N for instance.

\section{Acknowledgments}

We acknowledge Nathan Hara for his help on the l1-periodogram and the referee Etienne Artigau for his precious help regarding the improvement of the manuscript. The authors acknowledge the financial support of the SNSF. M.C. acknowledges the SNSF support under the Post-Doc Mobility grant P500PT\_211024. This project has received funding from the European Research Council (ERC) under the European Union’s Horizon 2020 research and innovation programme (grant agreement SCORE No 851555). This work has been carried out within the framework of the National Centre of Competence in Research PlanetS supported by the Swiss National Science Foundation under grants 51NF40\_182901 and 51NF40\_205606. 

This research has made use of NASA's Astrophysics Data System (ADS) bibliographic services. 
We acknowledge the community efforts devoted to the development of the following open-source packages that were used in this work: numpy (\href{http:\\numpy.org}{numpy.org}), matplotlib (\href{http:\\matplotlib.org}{matplotlib.org}), astropy (\href{http:\\astropy.org}{astropy.org}) and scikit-learn (\href{http:\\scikit-learn.org}{scikit-learn.org}).

\bibliographystyle{aa}
\bibliography{main}

\begin{appendix}

\section{Modified leave-p-out algorithm}
\label{appendix:a}

\begin{figure}[h!]
	
\centering
\includegraphics[width=9cm]{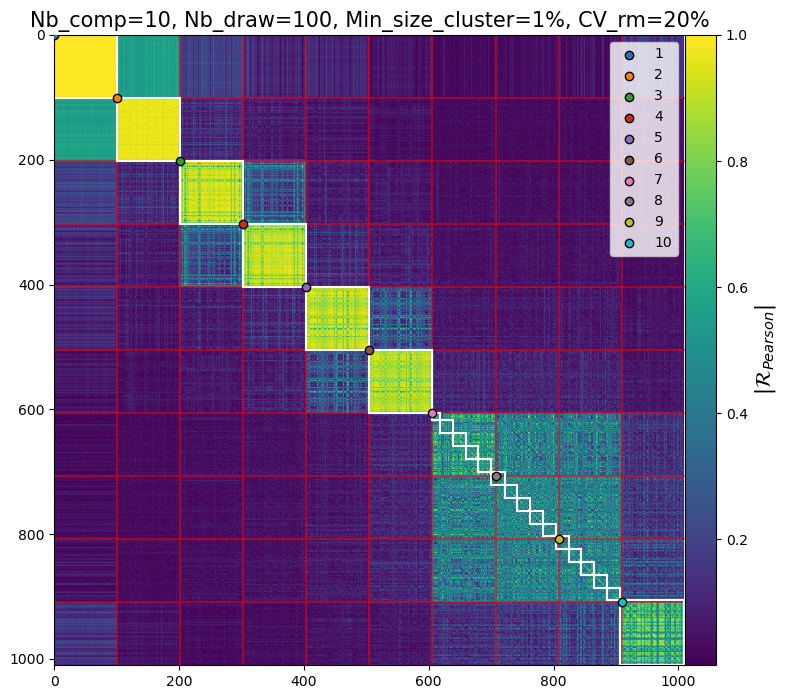}
\caption{Reproduction of the Fig. B.2 of Cretignier et al. (2022) for HD10700. Correlation matrix between the 1000 components obtained from the leave-20\%-out algorithm. Only 6 components satisfy the $med(R)>0.6$ criterion which is also the number of cluster that would have been fixed by eye.}
\label{FigCluster}
	
\end{figure}

In \citet{Cretignier(2022)}, the authors tried to assess the significance of the 10 first PCA components by randomly rejecting $p$ percent ($p=10\%$) of the dataset $N$ times. The algorithm then consisted to form clusters of the $10 \cdot N$ vectors before applying a criterion on the cluster size to establish the significance of a component. We slightly modified the code since we discovered that the algorithm was unstable numerically, in particular during the ordering step. Instead of ordering the components, we assumed that the $n^{th}$ component of any random realisation can be compared with another realisation. Say differently, the algorithm is not trying any ordering of the components to form the cluster since it assumes that the order of the components remains unchanged from one realisation to the other which will be the case for significant components. Moreover, we also removed the previous border detection algorithm and simply assumed that the size of the cluster is $N$ by construction. We reproduced the Fig.~B.2 of \citet{Cretignier(2022)} for HD10700 with the new algorithm in Fig.~\ref{FigCluster}. By eye, it can be noticed that 6 components are significant, which is similar to the 5 components found in the previous study. However, the correlation matrix is far cleaner in the present version. 

We then selected only components for which the median Pearson coefficient was higher than $med(\mathcal{R})=0.60$. We note that recently \citet{Ould(2023)} used a similar criterion with a higher threshold ($med(\mathcal{R})=0.95$). In fact, the authors likely misinterpreted in our previous paper the size of the cluster (fixed to be 95\% of the expected size $N$) with this value. Based on the targets we processed, the threshold at $med(\mathcal{R})=0.60$ was more useful to disentangle components from noise. As a detail, we slightly increased the fraction $p$ of data removed from 10\% to 20\% for the present paper.

\section{Detection of a new interference pattern on HARPS}
\label{appendix:b}

\begin{figure*}[h!]

    \centering
	\includegraphics[width=18cm]{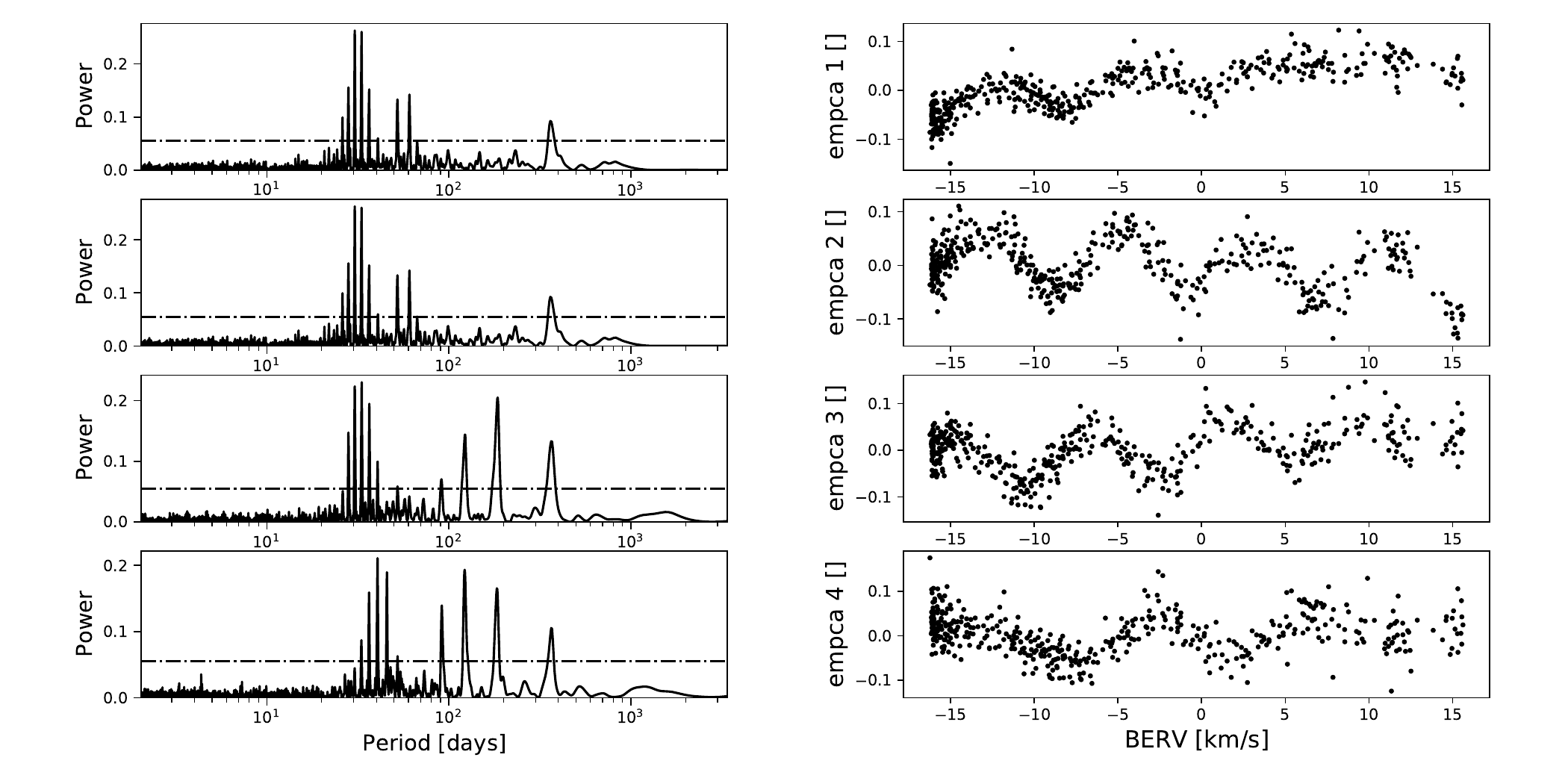}
	\caption{The four PCs vectors obtained by the PCA on the LBL RVs of HD20794. Even if the periodograms indicate a power excess around 32 days (left panels), the systematics is in fact related to a comb-like structure (see also Fig.\ref{FigPattern}) that is fixed in the terrestrial rest-frame. This is illustrated by plotting the PCA components as a function of the BERV (right panels). This comb-like structure, that is fixed in the laboratory rest-frame, then moves relatively to the stellar rest-frame with a 1-year periodicity due to the own Earth's revolution.}
	\label{FigColor3}
	
\end{figure*}

By performing a PCA directly on LBL RVs rather than on CBC RVs, we found out that several brightest stars where presenting in the PCs a forest of peaks around 30 days. An example is displayed in Fig.~\ref{FigColor3} for HD20794.Interesting enough, this effect is not due a $\sim$32-day systematic, but rather by a comb-like systematic moving relatively to the stellar rest-frame with a 1-year periodicity. This is visible in the figure where the PCs have been plotted as a function of the Barycentric Earth RV (BERV) and clearly exhibit an interference pattern structure. The leaking of power elsewhere than 1-year is quite common when a 1-year systematics is not randomly distributed in wavelength, but present some periodic structure as well. This is for instance also the case for the telluric oxygen lines, that are producing a forest of peaks at period around $\sim50$ days since they are produced in bands.

An interference pattern is the worst that can be encountered when working with PCA in LBL RVs. Indeed, since each stellar line is affected by a different phase of the signal, a perfectly regular interference pattern will take the shape of an hyperplan (2 components) in the LBL RVs space. If the pattern is irregular (in amplitude or frequency), its signatures can easily leak on several other components, explaining why the component is visible here on at least 4 PCs. Moreover, interference patterns produce usually large effects on individual lines (thus a large variance and are easily detected by PCA on LBL RVs) but since the phase of the signal is different for each line, the mean-effect is often negligible.

The new pattern found on HARPS is irregular in wavelength (appears and disappears) and is only clearly notable in the blue between 4150 and 4250 $\AA$. This pattern was likely induced by the old circular fibres on HARPS since such a pattern is no more visible on the HARPS data using the new octagonal fibres. The exact physical reason for the pattern is not yet fully understood. The periodicity of the pattern is 0.1 $\AA$
and has a peak-to-peak amplitude of 0.1\% which is close to the intrinsic precision that can be reached on an individual spectrum. A periodicity of 0.1 \ang{} is critical for LBL RVs since such wavelength scale is close to the width of the stellar lines which is in one hand maximising the effect of the pattern on an individual stellar line and in the other hand producing random phase from line-to-line making difficult line averaging method to highlight it. The fact that we are sensitive to so small systematic is however a good sign than any other systematics have been well corrected by YARARA. 

\begin{figure}[h!]
	\centering
	\includegraphics[width=9cm]{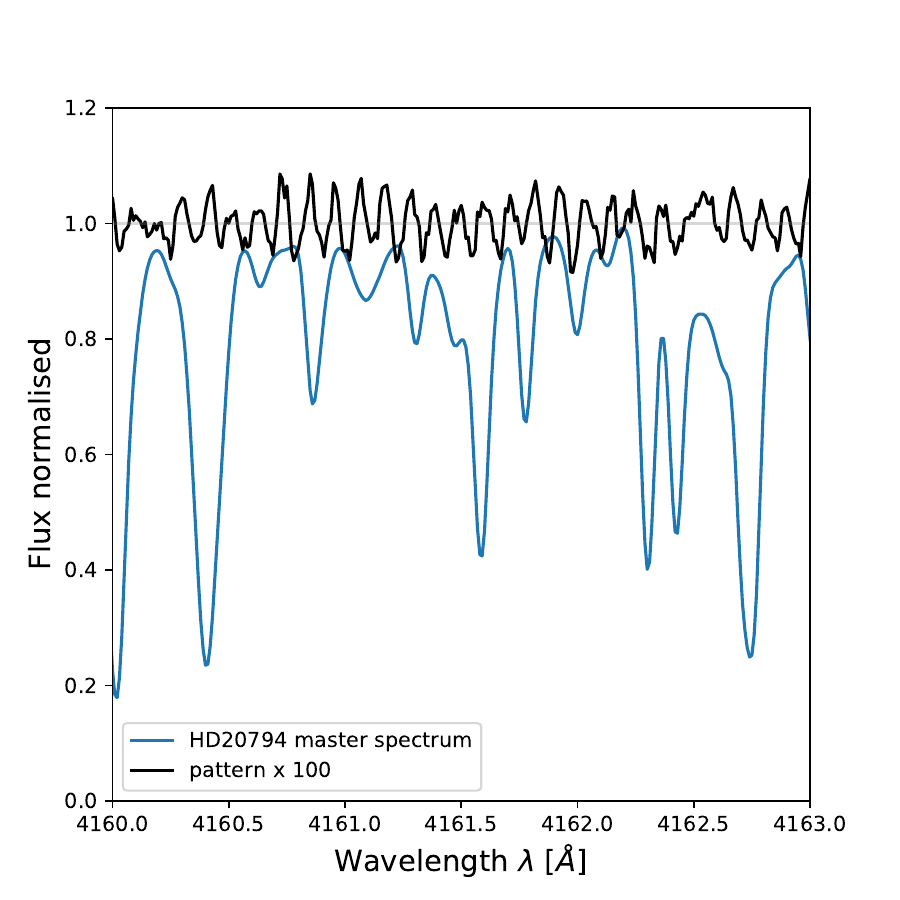}
	\caption{Extraction of a new systematic on HARPS by stacking in the terrestrial rest-frame the residuals spectra time-series YARARA-corrected of HD20794. The interference pattern present a periodicity of 0.1 \ang{} and a peak-to-peak amplitude of 0.1\% in flux units (magnified here by 100 to be visible). We displayed as a comparison the master spectrum of HD20794.}
	\label{FigPattern}
	
\end{figure}

We can wonder if such interference pattern could not be corrected at the spectrum level as we did in \citet{Cretignier(2021)} for another similar systematic. Unfortunately, this is not the case here due to the small amplitude of the pattern which cannot be detected on individual spectra. In order to be visible, we stacked in the terrestrial rest-frame the residual spectra time-series YARARA-corrected of HD20794, producing the vector visible in Fig.~\ref{FigPattern}. 

Despite all our attempts, we never managed to correct sufficiently the systematic such that this component was no more visible by PCA on LBL RVs. We therefore decided to not correct for it at all and opted for the CBC strategy presented in Sect.~\ref{sec:PCA}. This pattern is a good example of the kind of systematics that could blur the interpretation of PCs. As a final remark, we noted that the same pattern was not visible on HARPS after the fiber upgrade, but because the stars observed after fiber upgrade possess a smaller number of nightly observations, it is hard to conclude that the systematic has really disappeared.

\end{appendix}

\end{document}